\documentstyle[12pt]{article}
\begin{document}
\title{Non-Noether symmetries in Hamiltonian Dynamical Systems}
\author{George Chavchanidze} \date{} \maketitle
\thanks{Department of Theoretical Physics, A. Razmadze Institute of Mathematics, 1 Aleksidze Street, Tbilisi 0193, Georgia}
\begin{abstract}{\bf Abstract.} We discuss geometric properties of non-Noether symmetries and their possible applications in integrable Hamiltonian systems. Correspondence between non-Noether symmetries and conservation laws is revisited. It is shown that in regular Hamiltonian systems such a symmetries canonically lead to a Lax pairs on the algebra of linear operators on cotangent bundle over the phase space. Relationship between the non-Noether symmetries and other wide spread geometric methods of generating conservation laws such as bi-Hamiltonian formalism, bidifferential calculi and Fr\"{o}licher-Nijenhuis geometry is considered. It is proved that the integrals of motion associated with the continuous non-Noether symmetry are in involution whenever the generator of the symmetry satisfies a certain Yang-Baxter type equation. Action of one-parameter group of symmetry on algebra of integrals of motion is studied and involutivity of group orbits is discussed. Hidden non-Noether symmetries of Toda chain, nonlinear Schr\"{o}dinger equation, Korteweg-de Vries equations, Benney system, nonlinear water wave equations and Broer-Kaup system are revealed and discussed. \end{abstract}
\section{Introduction}
Symmetries play essential role in dynamical systems, because they usually simplify analysis of evolution equations and often provide quite elegant solution of problems that otherwise would be difficult to handle. In Lagrangian and Hamiltonian dynamical systems special role is played by Noether symmetries --- important class of symmetries that leave action invariant and have some exceptional features. In particular, Noether symmetries deserved special attention due to celebrated Noether's theorem, that established correspondence between symmetries, that leave action functional invariant, and conservation laws of Euler-Lagrange equations. This correspondence can be extended to Hamiltonian systems where it becomes tighter and more evident then in Lagrangian case and gives rise to Lie algebra homomorphism between Lie algebra of Noether symmetries and algebra of conservation laws (that form Lie algebra under Poisson bracket). \\
Role of symmetries that are not of Noether type was suppressed for quite a long time. However after some publications of Hojman, Harleston, Lutzky and others (see \cite{r16}, \cite{r36}, \cite{r39}, \cite{r40}, \cite{r49}-\cite{r54}) it became clear that non-Noether symmetries also can play important role in Lagrangian and Hamiltonian dynamics. In particular according to Lutzky \cite{r51}, in Lagrangian dynamics there is definite correspondence between non-Noether symmetries and conservation laws. Moreover, unlike Noether's case, each generator of non-Noether symmetry may produce whole family of conservation laws (maximal number of conservation laws that can be associated with non-Noether symmetry via Lutzky's theorem is equal to the dimension of configuration space of Lagrangian system). This fact makes non-Noether symmetries especially valuable in infinite dimensional dynamical systems, where potentially one can recover infinite sequence of conservation laws knowing single generator of non-Noether symmetry. \\
Existence of correspondence between non-Noether symmetries and conserved quantities, raised many questions concerning relationship among this type of symmetries and other geometric structures emerging in theory of integrable models. In particular one could notice suspicious similarity between the method of constructing conservation laws from generator of non-Noether symmetry and the way conserved quantities are produced in either Lax theory, bi-Hamiltonian formalism, bicomplex approach or Lenard scheme. It also raised natural question, whether set of conservation laws associated with non-Noether symmetry is involutive or not, and since it appeared that in general it may not be involutive, the need of involutivity criteria, similar to Yang-Baxter equation used in Lax theory or compatibility condition in bi-Hamiltonian formalism and bicomplex approach, emerged. It was also unclear how to construct conservation laws in case of infinite dimensional dynamical systems where volume forms used in Lutzky's construction are no longer well defined. Some of these questions were addressed in papers \cite{r11}-\cite{r14}, while in the present review we would like to summarize all these issues and to provide some samples of integrable models that possess non-Noether symmetries. \\
Review is organized as follows. In first section we briefly recall some aspects of geometric formulation of Hamiltonian dynamics. Further, in second section, correspondence between non-Noether symmetries and integrals of motion in regular Hamiltonian systems is discussed. Lutzky's theorem is reformulated in terms of bivector fields and alternative derivation of conserved quantities suitable for computations in infinite dimensional Hamiltonian dynamical systems is suggested. Non-Noether symmetries of two and three particle Toda chains are used to illustrate general theory. In the subsequent section geometric formulation of Hojman's theorem \cite{r36} is revisited and some samples are provided. Section 4 reveals correspondence between non-Noether symmetries and Lax pairs. It is shown that non-Noether symmetry canonically gives rise to a Lax pair of certain type. Lax pair is explicitly constructed in terms of Poisson bivector field and generator of symmetry. Sample of Toda chains are discussed. Next section deals with integrability issues. Analog of Yang-Baxter equation that, being satisfied by generator of symmetry, ensures involutivity of set of conservation laws produced by this symmetry, is introduced. Relationship between non-Noether symmetries and bi-Hamiltonian systems is considered in section 6. It is proved that under certain conditions, non-Noether symmetry endows phase space of regular Hamiltonian system with bi-Hamiltonian structure. We also discuss conditions under which non-Noether symmetry can be "recovered" from bi-Hamiltonian structure. Theory is illustrated by sample of Toda chains. Next section is devoted to bicomplexes and their relationship with non-Noether symmetries. Special kind of deformation of De Rham complex induced by symmetry is constructed in terms of Poisson bivector field and generator of symmetry. Samples of two and three particle Toda chain are discussed. Section 8 deals with Fr\"{o}licher-Nijenhuis recursion operators. It is shown that under certain condition non-Noether symmetry gives rise to invariant Fr\"{o}licher-Nijenhuis operator on tangent bundle over phase space. The last section of theoretical part contains some remarks on action of one-parameter group of symmetry on algebra of integrals of motion. Special attention is devoted to involutivity of group orbits. \\
Subsequent sections of present review provide samples of integrable models that possess interesting non-Noether symmetries. In particular section 10 reveals non-Noether symmetry of $n$-particle Toda chain. Bi-Hamiltonian structure, conservation laws, bicomplex, Lax pair and Fr\"{o}licher-Nijenhuis recursion operator of Toda hierarchy are constructed using this symmetry. Further we focus on infinite dimensional integrable Hamiltonian systems emerging in mathematical physics. In section 11 case of nonlinear Schr\"{o}dinger equation is discussed. Symmetry of this equation is identified and used in construction of involutive infinite sequence of conservation laws and bi-Hamiltonian structure of nonlinear Schr\"{o}dinger hierarchy. Section 12 deals with Korteweg-de Vries and modified Korteweg-de Vries equations. Non-Noether symmetries of these equations produce infinite number of conserved quantities in involution. The same symmetries give rise to bi-Hamiltonian structure of KdV hierarchies. Next section is devoted to non-Noether symmetries of integrable systems of nonlinear water wave equations, such as dispersive water wave system, Broer-Kaup system and dispersiveless long wave system. Last section focuses on Benney system and its non-Noether symmetry, that appears to be local, gives rise to infinite sequence of conserved densities of Benney hierarchy and endows it with bi-Hamiltonian structure. \\
\section{Regular Hamiltonian systems}
The basic concept in geometric formulation of Hamiltonian dynamics is notion of symplectic manifold. Such a manifold plays the role of the phase space of the dynamical system and therefore many properties of the dynamical system can be quite effectively investigated in the framework of symplectic geometry. Before we consider symmetries of the Hamiltonian dynamical systems, let us briefly recall some basic notions from symplectic geometry.\\
The symplectic manifold is a pair $(M, \omega )$ where $M$ is smooth even dimensional manifold and $\omega $ is a closed 

\begin{eqnarray}
\label{eq:e1}
d\omega = 0
\end{eqnarray}
and nondegenerate 2-form on $M$. Being nondegenerate means that contraction of arbitrary non-zero vector field with $\omega $ does not vanish 

\begin{eqnarray}
\label{eq:e2}
i_{X}\omega = 0 \Leftrightarrow X = 0
\end{eqnarray}
(here $i_{X}$ denotes contraction of the vector field $X$ with differential form). Otherwise one can say that $\omega $ is nondegenerate if its n-th outer power does not vanish ($\omega ^{n} \neq 0$) anywhere on $M$. In Hamiltonian dynamics $M$ is usually phase space of classical dynamical system with finite numbers of degrees of freedom and the symplectic form $\omega $ is basic object that defines Poisson bracket structure, algebra of Hamiltonian vector fields and the form of Hamilton's equations.\\
The symplectic form $\omega $ naturally defines isomorphism between vector fields and differential 1-forms on $M$ (in other words tangent bundle $TM$ of symplectic manifold can be quite naturally identified with cotangent bundle $T^{*}M$). The isomorphic map $\Phi _{\omega }$ from $TM$ into $T^{*}M$ is obtained by taking contraction of the vector field with $\omega $ 

\begin{eqnarray}
\label{eq:e3}
\Phi _{\omega }: X \rightarrow - i_{X}\omega 
\end{eqnarray}
(minus sign is the matter of convention). This isomorphism gives rise to natural classification of vector fields. Namely, vector field $X_{h}$ is said to be Hamiltonian if its image is exact 1-form or in other words if it satisfies Hamilton's equation 

\begin{eqnarray}
\label{eq:e4}
i_{X_{h}}\omega + dh = 0
\end{eqnarray}
for some function $h$ on $M$. Similarly, vector field $X$ is called locally Hamiltonian if it's image is closed 1-form 
\begin{eqnarray}
i_{X}\omega + u = 0, ~~~~~ du = 0
\end{eqnarray}
\\
One of the nice features of locally Hamiltonian vector fields, known as Liouville's theorem, is that these vector fields preserve symplectic form $\omega $. In other words Lie derivative of the symplectic form $\omega $ along arbitrary locally Hamiltonian vector field vanishes 
\begin{eqnarray}
L_{X}\omega = 0 \Leftrightarrow i_{X}\omega + du = 0, ~~~~~ du = 0
\end{eqnarray}
Indeed, using Cartan's formula that expresses Lie derivative in terms of contraction and exterior derivative 
\begin{eqnarray}
L_{X} = i_{X}d + di_{X}
\end{eqnarray}
one gets 
\begin{eqnarray}
L_{X}\omega = i_{X}d\omega + di_{X}\omega = di_{X}\omega 
\end{eqnarray}
(since $d\omega = 0$) but according to the definition of locally Hamiltonian vector field 
\begin{eqnarray}
di_{X}\omega = - du = 0
\end{eqnarray}
So locally Hamiltonian vector fields preserve $\omega $ and vise versa, if vector field preserves symplectic form $\omega $ then it is locally Hamiltonian.\\
Clearly, Hamiltonian vector fields constitute subset of locally Hamiltonian ones since every exact 1-form is also closed. Moreover one can notice that Hamiltonian vector fields form ideal in algebra of locally Hamiltonian vector fields. This fact can be observed as follows. First of all for arbitrary couple of locally Hamiltonian vector fields $X, Y$ we have $L_{X}\omega = L_{Y}\omega = 0$ and 
\begin{eqnarray}
L_{X}L_{Y}\omega - L_{Y}L_{X}\omega = L_{[X , Y]}\omega = 0
\end{eqnarray}
so locally Hamiltonian vector fields form Lie algebra (corresponding Lie bracket is ordinary commutator of vector fields). Further it is clear that for arbitrary Hamiltonian vector field $X_{h}$ and locally Hamiltonian one $Z$ one has 
\begin{eqnarray}
L_{Z}\omega = 0
\end{eqnarray}
and 
\begin{eqnarray}
i_{X_{h}}\omega + dh = 0
\end{eqnarray}
that implies 
\begin{eqnarray}
L_{Z}(i_{X_{h}}\omega + dh) = L_{[Z , X_{h}]}\omega + i_{X_{h}}L_{Z}\omega + dL_{Z}h = L_{[Z , X_{h}]}\omega + dL_{Z}h = 0
\end{eqnarray}
thus commutator $[Z , X_{h}]$ is Hamiltonian vector field $X_{L_{Z}h}$, or in other words Hamiltonian vector fields form ideal in algebra of locally Hamiltonian vector fields.\\
Isomorphism $\Phi _{\omega }$ can be extended to higher order vector fields and differential forms by linearity and multiplicativity. Namely, 
\begin{eqnarray}
\Phi _{\omega }(X \wedge Y) = \Phi _{\omega }(X) \wedge \Phi _{\omega }(Y)
\end{eqnarray}
Since $\Phi _{\omega }$ is isomorphism, the symplectic form $\omega $ has unique counter image $W$ known as Poisson bivector field. Property $d\omega = 0$ together with non degeneracy implies that bivector field $W$ is also nondegenerate ($W^{n} \neq 0$) and satisfies condition 

\begin{eqnarray}
\label{eq:e5}
[W , W] = 0 
\end{eqnarray}
where bracket $[ , ]$ known as Schouten bracket or supercommutator, is actually graded extension of ordinary commutator of vector fields to the case of multivector fields, and can be defined by linearity and derivation property 
\begin{eqnarray}
[C_{1} \wedge C_{2} \wedge ... \wedge C_{n} , S_{1} \wedge S_{2} \wedge ... \wedge S_{n}] = \nonumber \\(- 1)^{p + q}[C_{p} , S_{q}] \wedge C_{1} \wedge C_{2} \wedge ... \wedge \hat{C}_{p} \wedge ... \wedge C_{n} \wedge S_{1} \wedge S_{2} \wedge ... \wedge \hat{S}_{q} \wedge ...\wedge S_{n} 
\end{eqnarray}
where over hat denotes omission of corresponding vector field. In terms of the bivector field $W$ Liouville's theorem mentioned above can be rewritten as follows 

\begin{eqnarray}
\label{eq:e6}
[W(u) , W] = 0 \Leftrightarrow du = 0
\end{eqnarray}
for each 1-form $u$. It follows from graded Jacoby identity satisfied by Schouten bracket and property $[W , W] = 0$ satisfied by Poisson bivector field. \\
Being counter image of symplectic form, $W$ gives rise to map $\Phi _{W}$, transforming differential 1-forms into vector fields, which is inverted to the map $\Phi _{\omega }$ and is defined by 
\begin{eqnarray}
\Phi _{W}: u \rightarrow W(u); ~~~~~ \Phi _{W}\Phi _{\omega } = id
\end{eqnarray}
Further we will often use these maps. \\
In Hamiltonian dynamical systems Poisson bivector field is geometric object that underlies definition of Poisson bracket --- kind of Lie bracket on algebra of smooth real functions on phase space. In terms of bivector field $W$ Poisson bracket is defined by 

\begin{eqnarray}
\label{eq:e7}
\{f , g\} = W(df \wedge dg)
\end{eqnarray}
The condition $[W , W] = 0$ satisfied by bivector field ensures that for every triple $(f, g, h)$ of smooth functions on the phase space the Jacobi identity 

\begin{eqnarray}
\label{eq:e8}
\{f\{g , h\}\} + \{h\{f , g\}\} + \{g\{h , f\}\} = 0.
\end{eqnarray}
is satisfied. Interesting property of the Poisson bracket is that map from algebra of real smooth functions on phase space into algebra of Hamiltonian vector fields, defined by Poisson bivector field 
\begin{eqnarray}
f \rightarrow X_{f} = W(df)
\end{eqnarray}
appears to be homomorphism of Lie algebras. In other words commutator of two vector fields associated with two arbitrary functions reproduces vector field associated with Poisson bracket of these functions 

\begin{eqnarray}
\label{eq:e9}
[X_{f} , X_{g}] = X_{\{f , g\}}
\end{eqnarray}
This property is consequence of the Liouville theorem and definition of Poisson bracket. Further we also need another useful property of Hamiltonian vector fields and Poisson bracket 

\begin{eqnarray}
\label{eq:e10}
\{f , g\} = W(df \wedge dg) = \omega (X_{f} \wedge X_{g}) = L_{X_{f}}g = - L_{X_{g}}g
\end{eqnarray}
it also follows from Liouville theorem and definition of Hamiltonian vector fields and Poisson brackets. \\
To define dynamics on $M$ one has to specify time evolution of observables (smooth functions on $M$). In Hamiltonian dynamical systems time evolution is governed by Hamilton's equation 

\begin{eqnarray}
\label{eq:e11}
\frac{ d }{dt }f = \{h , f\}
\end{eqnarray}
where $h$ is some fixed smooth function on the phase space called Hamiltonian. In local coordinate frame $z_{k}$ bivector field $W$ has the form 
\begin{eqnarray}
W = W_{bc} \frac{ \partial }{\partial z_{b} } \wedge \frac{ \partial }{\partial z_{c} }
\end{eqnarray}
and the Hamilton's equation rewritten in terms of local coordinates takes the form 
\begin{eqnarray}
\dot{z}_{b} = W_{bc} \frac{ \partial h }{\partial z_{c} }
\end{eqnarray}
Note that functions $W_{ab}$ are not arbitrary, to ensure validity of $[W , W] = 0$ condition they should fulfill restriction 
\begin{eqnarray}
\sum ^{ n }_{a = 1 } \left [ W_{ab} \frac{ \partial W_{cd} }{\partial z_{a} } + W_{ac} \frac{ \partial W_{bd} }{\partial z_{a} } + W_{ad} \frac{ \partial W_{bc} }{\partial z_{a} } \right ] = 0 
\end{eqnarray}
and in the same time determinant of matrix formed by functions $W_{ab}$ should not vanish to ensure that Poisson bivector field $W$ is nondegenerate. \\
\section{Non-Noether symmetries}
Now let us focus on symmetries of Hamilton's equation (\ref{eq:e11}). Generally speaking, symmetries play very important role in Hamiltonian dynamics due to different reasons. They not only give rise to conservation laws but also often provide very effective solutions to problems that otherwise would be difficult to solve. Here we consider special class of symmetries of Hamilton's equation called non-Noether symmetries. Such a symmetries appear to be closely related to many geometric concepts used in Hamiltonian dynamics including bi-Hamiltonian structures, Fr\"{o}licher-Nijenhuis operators, Lax pairs and bicomplexes.\\
Before we proceed let us recall that each vector field $E$ on the phase space generates the one-parameter continuous group of transformations $g_{a} = e^{aL_{E}}$ (here $L$ denotes Lie derivative) that acts on the observables as follows 

\begin{eqnarray}
\label{eq:e12}
g_{a}(f) = e^{aL_{E}}(f) = f + aL_{E}f + \frac{ 1 }{2 }(aL_{E})^{2}f + ...
\end{eqnarray}
Such a group of transformation is called symmetry of Hamilton's equation (\ref{eq:e11}) if it commutes with time evolution operator 

\begin{eqnarray}
\label{eq:e13}
\frac{ d }{dt } g_{a}(f) = g_{a}(\frac{ d }{dt }f) 
\end{eqnarray}
in terms of the vector fields this condition means that the generator $E$ of the group $g_{a}$ commutes with the vector field $W(h) = \{h , \}$, i. e. 

\begin{eqnarray}
\label{eq:e14}
[E , W(h)] = 0.
\end{eqnarray}
However we would like to consider more general case where $E$ is time dependent vector field on phase space. In this case (\ref{eq:e14}) should be replaced with 

\begin{eqnarray}
\label{eq:e15}
\frac{ \partial }{\partial t }E = [E , W(h)].
\end{eqnarray}
\\
Further one should distinguish between groups of symmetry transformations generated by Hamiltonian, locally Hamiltonian and non-Hamiltonian vector fields. First kind of symmetries are known as Noether symmetries and are widely used in Hamiltonian dynamics due to their tight connection with conservation laws. Second group of symmetries is less interesting but locally they also lead to conservation laws. While third group of symmetries that further will be referred as non-Noether symmetries seems to play important role in integrability issues due to their remarkable relationship with bi-Hamiltonian structures and Fr\"{o}licher-Nijenhuis operators. Thus if in addition to (\ref{eq:e14}) the vector field $E$ does not preserve Poisson bivector field $[E , W] \neq 0$ then $g_{a}$ is called non-Noether symmetry.\\
Now let us focus on non-Noether symmetries. We would like to show that the presence of such a symmetry essentially enriches the geometry of the phase space and under the certain conditions can ensure integrability of the dynamical system. Before we proceed let us recall that the non-Noether symmetry leads to a number of integrals of motion. More precisely the relationship between non-Noether symmetries and the conservation laws is described by the following theorem. This theorem was proposed by Lutzky in \cite{r51}. Here it is reformulated in terms of Poisson bivector field. \\
{\bf Theorem 1.} Let $(M , h)$ be regular Hamiltonian system on the $2n$-dimensional Poisson manifold $M$. Then, if the vector field $E$ generates non-Noether symmetry, the functions 

\begin{eqnarray}
\label{eq:e16}
Y^{(k)} = \frac{ \hat{W}^{k} \wedge W^{n - k} }{W^{n} } ~~~~~k = 1,2, ... n
\end{eqnarray}
where $\hat{W} = [E , W]$, are integrals of motion. \\
{\bf Proof:} By the definition 
\begin{eqnarray}
\hat{W}^{k} \wedge W^{n - k} = Y^{(k)}W^{n}.
\end{eqnarray}
(definition is correct since the space of $2n$ degree multivector fields on $2n$ degree manifold is one dimensional). Let us take time derivative of this expression along the vector field $W(h)$, 
\begin{eqnarray}
\frac{ d }{dt }\hat{W}^{k} \wedge W^{n - k} = (\frac{ d }{dt }Y^{(k)})W^{n} + Y^{(k)}[W(h) , W^{n}]
\end{eqnarray}
or 

\begin{eqnarray}
\label{eq:e17}
k(\frac{ d }{dt }\hat{W}) \wedge \hat{W}^{k - 1} \wedge W^{n - k} + (n - k)[W(h) , W] \wedge \hat{W}^{k} \wedge W^{n - k - 1} = \nonumber \\(\frac{ d }{dt }Y^{(k)})W^{n} + nY^{(k)}[W(h) , W] \wedge W^{n - 1}
\end{eqnarray}
but according to the Liouville theorem the Hamiltonian vector field preserves $W$ i. e. 
\begin{eqnarray}
\frac{ d }{dt }W = [W(h) , W] = 0
\end{eqnarray}
hence, by taking into account that 
\begin{eqnarray}
\frac{ d }{dt }E= \frac{ \partial }{\partial t }E + [W(h) , E] = 0
\end{eqnarray}
we get 
\begin{eqnarray}
\frac{ d }{dt }\hat{W} = \frac{ d }{dt }[E , W] = [\frac{ d }{dt }E , W] + [E[W(h) , W]] = 0.
\end{eqnarray}
and as a result (\ref{eq:e17}) yields 
\begin{eqnarray}
\frac{ d }{dt }Y^{(k)}W^{n} = 0
\end{eqnarray}
but since the dynamical system is regular ($W^{n} \neq 0$) we obtain that the functions $Y^{(k)}$ are integrals of motion. \\
{\bf Remark 1.} Instead of conserved quantities $Y^{(1)} ... Y^{(n)}$, the solutions $c_{1} ... c_{n}$ of the secular equation 

\begin{eqnarray}
\label{eq:e18}
(\hat{W} - cW)^{n} = 0
\end{eqnarray}
can be associated with the generator of symmetry. By expanding expression (\ref{eq:e18}) it is easy to verify that the conservation laws $Y^{(k)}$ can be expressed in terms of the integrals of motion $c_{1} ... c_{n}$ in the following way 

\begin{eqnarray}
\label{eq:e19}
Y^{(k)} = \frac{ (n - k)! k! }{n! } \sum _{i_{1}<i_{2}<...<i_{k}} c_{i_{1}}c_{i_{2}} ... c_{i_{k}}
\end{eqnarray}
Note also that conservation laws $Y^{(k)}$ can be also defined by means of symplectic form $\omega $ using the following formula 

\begin{eqnarray}
\label{eq:e20}
Y^{(k)} = \frac{ (L_{E}\omega )^{k} \wedge \omega ^{n - k} }{\omega ^{n} } ~~~~~k = 1,2, ... n
\end{eqnarray}
while $c_{1} ... c_{n}$ conservation laws can be derived from the secular equation 

\begin{eqnarray}
\label{eq:e21}
(L_{E}\omega - c\omega )^{n} = 0
\end{eqnarray}
However all these expressions fail in case of infinite dimensional Hamiltonian systems where the volume form 
\begin{eqnarray}
\Omega = \omega ^{n}
\end{eqnarray}
does not exist since $n = \infty $. But fortunately in these case one can define conservation laws using alternative formula 

\begin{eqnarray}
\label{eq:e22}
C^{(k)} = i_{W^{k}}(L_{E}\omega )^{k}
\end{eqnarray}
as far as it involves only finite degree differential forms $(L_{E}\omega )^{k}$ and well defined multivector fields $W^{k}$, further we will use this expression in construction of infinite sequence of conservation laws in Korteweg-De Vries, modified Korteweg-De Vries and nonlinear Schr\"{o}dinger equations. Note that in finite dimensional case the sequence of conservation laws $C^{(k)}$ is related to families of conservation laws $Y^{(k)}$ and $c_{k}$ in the following way 

\begin{eqnarray}
\label{eq:e23}
C^{(k)} = \sum _{i_{1}<i_{2}<...<i_{k}} c_{i_{1}}c_{i_{2}} ... c_{i_{k}} = \frac{ n! }{(n - k)! k! } Y^{(k)} 
\end{eqnarray}
Note also that by taking Lie derivative of known conservation along the generator of symmetry $E$ one can construct new conservation laws 
\begin{eqnarray}
\frac{ d }{dt }Y = L_{X_{h}}Y = 0 \Rightarrow \frac{ d }{dt }L_{E}Y = L_{X_{h}}L_{E}Y = L_{E}L_{X_{h}}Y = 0
\end{eqnarray}
since $[E , X_{h}] = 0$. \\
{\bf Remark 2.} Besides continuous non-Noether symmetries generated by non-Hamiltonian vector fields one may encounter discrete non-Noether symmetries --- noncannonical transformations that doesn't necessarily form group but commute with evolution operator 
\begin{eqnarray}
\frac{ d }{dt } g(f) = g(\frac{ d }{dt }f) 
\end{eqnarray}
Such a symmetries give rise to the same conservation laws 

\begin{eqnarray}
\label{eq:e24}
Y^{(k)} = \frac{ g(W)^{k} \wedge W^{n - k} }{W^{n} } ~~~~~k = 1,2, ... n
\end{eqnarray}
\\
{\bf Sample.} Let $M$ be $R^{4}$ with coordinates $z_{1}, z_{2}, z_{3}, z_{4}$ and Poisson bivector field 

\begin{eqnarray}
\label{eq:e25}
W = \frac{ \partial }{\partial z_{1} } \wedge \frac{ \partial }{\partial z_{3} } + \frac{ \partial }{\partial z_{2} } \wedge \frac{ \partial }{\partial z_{4} } 
\end{eqnarray}
and let's take 

\begin{eqnarray}
\label{eq:e26}
h = \frac{ 1 }{2 }z_{1}^{2} + \frac{ 1 }{2 }z_{2}^{2} + e^{z_{3} - z_{4}} 
\end{eqnarray}
This is so called two particle non periodic Toda model. One can check that the vector field 
\begin{eqnarray}
E = \sum ^{ 4 }_{a = 1 } E_{a} \frac{ \partial }{\partial z_{a} } 
\end{eqnarray}
with components 

\begin{eqnarray}
\label{eq:e27}
E_{1} = \frac{ 1 }{2 }z_{1}^{2} - e^{z_{3} - z_{4}} - \frac{ t }{2 }(z_{1} + z_{2})e^{z_{3} - z_{4}}\nonumber \\E_{2} = \frac{ 1 }{2 }z_{2}^{2} + 2e^{z_{3} - z_{4}} + \frac{ t }{2 }(z_{1} + z_{2})e^{z_{3} - z_{4}}\nonumber \\E_{3} = 2z_{1} + \frac{ 1 }{2 }z_{2} + \frac{ t }{2 }(z_{1}^{2} + e^{z_{3} - z_{4}})\nonumber \\E_{4} = z_{2} - \frac{ 1 }{2 }z_{1} + \frac{ t }{2 }(z_{2}^{2} + e^{z_{3} - z_{4}}) 
\end{eqnarray}
satisfies (\ref{eq:e15}) condition and as a result generates symmetry of the dynamical system. The symmetry appears to be non-Noether with Schouten bracket $[E , W]$ equal to 

\begin{eqnarray}
\label{eq:e28}
\hat{W} = [E , W] = z_{1} \frac{ \partial }{\partial z_{1} } \wedge \frac{ \partial }{\partial z_{3} } + z_{2} \frac{ \partial }{\partial z_{2} } \wedge \frac{ \partial }{\partial z_{4} } + \nonumber \\e^{z_{3} - z_{4}} \frac{ \partial }{\partial z_{1} } \wedge \frac{ \partial }{\partial z_{2} } + \frac{ \partial }{\partial z_{3} } \wedge \frac{ \partial }{\partial z_{4} } 
\end{eqnarray}
calculating volume vector fields $\hat{W}^{k} \wedge W^{n - k}$ gives rise to 
\begin{eqnarray}
W \wedge W = - 2 \frac{ \partial }{\partial z_{1} } \wedge \frac{ \partial }{\partial z_{2} } \wedge \frac{ \partial }{\partial z_{3} } \wedge \frac{ \partial }{\partial z_{4} }\nonumber \\\hat{W} \wedge W = - (z_{1} + z_{2}) \frac{ \partial }{\partial z_{1} } \wedge \frac{ \partial }{\partial z_{2} } \wedge \frac{ \partial }{\partial z_{3} } \wedge \frac{ \partial }{\partial z_{4} }\nonumber \\\hat{W} \wedge \hat{W} = - 2(z_{1}z_{2} - e^{z_{3} - z_{4}}) \frac{ \partial }{\partial z_{1} } \wedge \frac{ \partial }{\partial z_{2} } \wedge \frac{ \partial }{\partial z_{3} } \wedge \frac{ \partial }{\partial z_{4} } 
\end{eqnarray}
and the conservation laws associated with this symmetry are just 

\begin{eqnarray}
\label{eq:e29}
Y^{(1)} = \frac{ \hat{W} \wedge W }{W \wedge W } = \frac{ 1 }{2 }(z_{1} + z_{2})\nonumber \\Y^{(2)} = \frac{ \hat{W} \wedge \hat{W} }{W \wedge W } = z_{1}z_{2} - e^{z_{3} - z_{4}}
\end{eqnarray}
It is remarkable that the same symmetry is also present in higher dimensions. For example in case when $M$ is $R^{6}$ with coordinates 
\begin{eqnarray}
z_{1}, z_{2}, z_{3}, z_{4}, z_{5}, z_{6}
\end{eqnarray}
Poisson bivector equal to 

\begin{eqnarray}
\label{eq:e30}
W = \frac{ \partial }{\partial z_{1} } \wedge \frac{ \partial }{\partial z_{4} } + \frac{ \partial }{\partial z_{2} } \wedge \frac{ \partial }{\partial z_{5} } + \frac{ \partial }{\partial z_{3} } \wedge \frac{ \partial }{\partial z_{6} } 
\end{eqnarray}
and the following Hamiltonian 

\begin{eqnarray}
\label{eq:e31}
h = \frac{ 1 }{2 }z_{1}^{2} + \frac{ 1 }{2 }z_{2}^{2} + \frac{ 1 }{2 }z_{3}^{2} + e^{z_{4} - z_{5}} + e^{z_{5} - z_{6}} 
\end{eqnarray}
we still can construct symmetry similar to (\ref{eq:e27}). More precisely the vector field 
\begin{eqnarray}
E = \sum ^{ 6 }_{a = 1 } E_{a} \frac{ \partial }{\partial z_{a} } 
\end{eqnarray}
with components specified as follows 

\begin{eqnarray}
\label{eq:e32}
E_{1} = \frac{ 1 }{2 }z_{1}^{2} - 2e^{z_{4} - z_{5}} - \frac{ t }{2 }(z_{1} + z_{2})e^{z_{4} - z_{5}}\nonumber \\E_{2} = \frac{ 1 }{2 }z_{2}^{2} + 3e^{z_{4} - z_{5}} - e^{z_{5} - z_{6}} + \frac{ t }{2 }(z_{1} + z_{2})e^{z_{4} - z_{5}}\nonumber \\E_{3} = \frac{ 1 }{2 }z_{3}^{2} + 2e^{z_{5} - z_{6}} + \frac{ t }{2 }(z_{2} + z_{3})e^{z_{5} - z_{6}} 
\end{eqnarray}
\begin{eqnarray}
E_{4} = 3z_{1} + \frac{ 1 }{2 }z_{2} + \frac{ 1 }{2 }z_{3} + \frac{ t }{2 }(z_{1}^{2} + e^{z_{4} - z_{5}})\nonumber \\E_{5} = 2z_{2} - \frac{ 1 }{2 }z_{1} + \frac{ 1 }{2 }z_{3} + \frac{ t }{2 }(z_{2}^{2} + e^{z_{4} - z_{5}} + e^{z_{5} - z_{6}})\nonumber \\E_{6} = z_{3} - \frac{ 1 }{2 }z_{1} - \frac{ 1 }{2 }z_{2} + \frac{ t }{2 }(z_{3}^{2} + e^{z_{5} - z_{6}}) 
\end{eqnarray}
satisfies (\ref{eq:e15}) condition and generates non-Noether symmetry of the dynamical system (three particle non periodic Toda chain). Calculating Schouten bracket $[E , W]$ gives rise to expression 

\begin{eqnarray}
\label{eq:e33}
\hat{W} = [E , W] = z_{1} \frac{ \partial }{\partial z_{1} } \wedge \frac{ \partial }{\partial z_{4} } + z_{2} \frac{ \partial }{\partial z_{2} } \wedge \frac{ \partial }{\partial z_{5} } + z_{3} \frac{ \partial }{\partial z_{3} } \wedge \frac{ \partial }{\partial z_{6} } +\nonumber \\e^{z_{4} - z_{5}} \frac{ \partial }{\partial z_{1} } \wedge \frac{ \partial }{\partial z_{2} } + e^{z_{5} - z_{6}} \frac{ \partial }{\partial z_{2} } \wedge \frac{ \partial }{\partial z_{3} } + \nonumber \\\frac{ \partial }{\partial z_{3} } \wedge \frac{ \partial }{\partial z_{4} } + \frac{ \partial }{\partial z_{4} } \wedge \frac{ \partial }{\partial z_{5} } + \frac{ \partial }{\partial z_{5} } \wedge \frac{ \partial }{\partial z_{6} } 
\end{eqnarray}
Volume multivector fields $\hat{W}^{k} \wedge W^{n - k}$ can be calculated in the manner similar to $R^{4}$ case and give rise to the well known conservation laws of three particle Toda chain. 

\begin{eqnarray}
\label{eq:e34}
Y^{(1)} = \frac{ 1 }{6 } (z_{1} + z_{2} + z_{3}) = \frac{ \hat{W} \wedge W \wedge W }{W \wedge W \wedge W } \nonumber \\Y^{(2)} = \frac{ 1 }{3 } (z_{1}z_{2} + z_{1}z_{3} + z_{2}z_{3} - e^{z_{4} - z_{5}} - e^{z_{5} - z_{6}}) = \frac{ \hat{W} \wedge \hat{W} \wedge W }{W \wedge W \wedge W } \nonumber \\Y^{(3)} = z_{1}z_{2}z_{3} - z_{3}e^{z_{4} - z_{5}} - z_{1}e^{z_{5} - z_{6}} = \frac{ \hat{W} \wedge \hat{W} \wedge \hat{W} }{W \wedge W \wedge W } 
\end{eqnarray}
\\
\section{Non-Liouville symmetries}
Besides Hamiltonian dynamical systems that admit invariant symplectic form $\omega $, there are dynamical systems that either are not Hamiltonian or admit Hamiltonian realization but explicit form of symplectic structure $\omega $ is unknown or too complex. However usually such a dynamical systems possess invariant volume form $\Omega $ which like symplectic form can be effectively used in construction of conservation laws. Note that volume form for given manifold is arbitrary differential form of maximal degree (equal to the dimension of manifold). In case of regular Hamiltonian systems, n-th outer power of the symplectic form $\omega $ naturally gives rise to the invariant volume form known as Liouville form 
\begin{eqnarray}
\Omega = \omega ^{n}
\end{eqnarray}
and sometimes it is easier to work with $\Omega $ rather then with symplectic form itself. In generic Liouville dynamical system time evolution is governed by equations of motion 

\begin{eqnarray}
\label{eq:e35}
\frac{ d }{dt }f = X(f) 
\end{eqnarray}
where $X$ is some smooth vector field that preserves Liouville volume form $\Omega $ 
\begin{eqnarray}
\frac{ d }{dt }\Omega = L_{X}\Omega = 0 
\end{eqnarray}
Symmetry of equations of motion still can be defined by condition 
\begin{eqnarray}
\frac{ d }{dt } g_{a}(f) = g_{a}(\frac{ d }{dt }f) 
\end{eqnarray}
that in terms of vector fields implies that generator of symmetry $E$ should commute with time evolution operator $X$ 
\begin{eqnarray}
[E , X] = 0
\end{eqnarray}
Throughout this chapter symmetry will be called non-Liouville if it is not conformal symmetry of $\Omega $, or in other words if 
\begin{eqnarray}
L_{E}\Omega \neq c\Omega 
\end{eqnarray}
for any constant $c$. Such a symmetries may be considered as analog of non-Noether symmetries defined in Hamiltonian systems and similarly to the Hamiltonian case one can try to construct conservation laws by means of generator of symmetry $E$ and invariant differential form $\Omega $. Namely we have the following theorem, which is reformulation of Hojman's theorem in terms of Liouville volume form. \\
{\bf Theorem 2.} Let $(M, X, \Omega )$ be Liouville dynamical system on the smooth manifold $M$. Then, if the vector field $E$ generates non-Liouville symmetry, the function 

\begin{eqnarray}
\label{eq:e36}
J = \frac{ L_{E}\Omega }{\Omega } 
\end{eqnarray}
is conservation law. \\
{\bf Proof:} By the definition 
\begin{eqnarray}
L_{E}\Omega = J\Omega .
\end{eqnarray}
and $J$ is not just constant (again definition is correct since the space of volume forms is one dimensional). By taking Lie derivative of this expression along the vector field $X$ that defines time evolution we get 
\begin{eqnarray}
L_{X}L_{E}\Omega = L_{[X , E]}\Omega + L_{E}L_{X}\Omega =\nonumber \\L_{X}(J\Omega ) = (L_{X}J)\Omega + JL_{X}\Omega 
\end{eqnarray}
but since Liouville volume form is invariant $L_{X}\Omega = 0$ and vector field $E$ is generator of symmetry satisfying $[E , X] = 0$ commutation relation we obtain 
\begin{eqnarray}
(L_{X}J)\Omega = 0
\end{eqnarray}
or 
\begin{eqnarray}
\frac{ d }{dt }J = L_{X}J = 0
\end{eqnarray}
\\
{\bf Remark 3.} In fact theorem is valid for larger class of symmetries. Namely one can consider symmetries with time dependent generators. Note however that in this case condition $[E , X] = 0$ should be replaced by 
\begin{eqnarray}
\frac{ \partial }{\partial t }E = [E , X]
\end{eqnarray}
Note also that by calculating Lie derivative of conservation law $J$ along generator of the symmetry $E$ one can recover additional conservation laws 
\begin{eqnarray}
J^{(m)} = (L_{E})^{m}\Omega 
\end{eqnarray}
\\
{\bf Sample.} Let us consider symmetry of three particle non periodic Toda chain. This dynamical system with equations of motion defined by the vector field 
\begin{eqnarray}
X = - e^{z_{4} - z_{5}} \frac{ \partial }{\partial z_{1} } + (e^{z_{4} - z_{5}} - e^{z_{5} - z_{6}}) \frac{ \partial }{\partial z_{2} } + e^{z_{5} - z_{6}} \frac{ \partial }{\partial z_{3} } \nonumber \\+ z_{1} \frac{ \partial }{\partial z_{4} } + z_{2} \frac{ \partial }{\partial z_{5} } + z_{3} \frac{ \partial }{\partial z_{6} } 
\end{eqnarray}
possesses invariant volume form 
\begin{eqnarray}
\Omega = dz_{1} \wedge dz_{2} \wedge dz_{3} \wedge dz_{4} \wedge dz_{5} \wedge dz_{6} 
\end{eqnarray}
One can check that $\Omega $ is really invariant volume form, i.e. Lie derivative of $\Omega $ along $X$ vanishes 
\begin{eqnarray}
\frac{ d }{dt }\Omega = L_{X}\Omega = \left [ \frac{ \partial X_{1} }{\partial z_{1} } + \frac{ \partial X_{2} }{\partial z_{2} } + \frac{ \partial X_{3} }{\partial z_{3} } + \frac{ \partial X_{4} }{\partial z_{4} } + \frac{ \partial X_{5} }{\partial z_{5} } + \frac{ \partial X_{6} }{\partial z_{6} } \right ] \Omega = 0 
\end{eqnarray}
The symmetry (\ref{eq:e32}) is clearly non-Liouville one as far as 
\begin{eqnarray}
L_{E}\Omega = \left [ \frac{ \partial E_{1} }{\partial z_{1} } + \frac{ \partial E_{2} }{\partial z_{2} } + \frac{ \partial E_{3} }{\partial z_{3} } + \frac{ \partial E_{4} }{\partial z_{4} } + \frac{ \partial E_{5} }{\partial z_{5} } + \frac{ \partial E_{6} }{\partial z_{6} } \right ] \Omega = \nonumber \\(z_{1} + z_{2} + z_{3})dz_{1} \wedge dz_{2} \wedge dz_{3} \wedge dz_{4} \wedge dz_{5} \wedge dz_{6} = \nonumber \\(z_{1} + z_{2} + z_{3}) \Omega 
\end{eqnarray}
and main conservation law associated with this symmetry via Theorem 2 is total momentum 
\begin{eqnarray}
J = \frac{ L_{E}\Omega }{\Omega } = z_{1} + z_{2} + z_{3} 
\end{eqnarray}
Other conservation laws can be recovered by taking Lie derivative of $J$ along generator of symmetry $E$, in particular 
\begin{eqnarray}
J^{(1)} = L_{E}J = \nonumber \\\frac{ 1 }{2 }z_{1}^{2} + \frac{ 1 }{2 }z_{2}^{2} + \frac{ 1 }{2 }z_{3}^{2} + e^{z_{4} - z_{5}} + e^{z_{5} - z_{6}}\nonumber \\J^{(2)} = L_{E}J^{(1)} = \nonumber \\\frac{ 1 }{2 } (z_{1}^{3} + z_{2}^{3} + z_{3}^{3}) + \frac{ 3 }{2 } (z_{1} + z_{2})e^{z_{4} - z_{5}} + \frac{ 3 }{2 } (z_{2} + z_{3})e^{z_{5} - z_{6}} 
\end{eqnarray}
\\
\section{Lax Pairs}
Presence of the non-Noether symmetry not only leads to a sequence of conservation laws, but also endows the phase space with a number of interesting geometric structures and it appears that such a symmetry is related to many important concepts used in theory of dynamical systems. One of the such concepts is Lax pair that plays quite important role in construction of completely integrable models. Let us recall that Lax pair of Hamiltonian system on Poisson manifold $M$ is a pair $(L , P)$ of smooth functions on $M$ with values in some Lie algebra $g$ such that the time evolution of $L$ is given by adjoint action 

\begin{eqnarray}
\label{eq:e37}
\frac{ d }{dt }L = [L , P] = - ad_{P}L 
\end{eqnarray}
where $[ , ]$ is a Lie bracket on $g$. It is well known that each Lax pair leads to a number of conservation laws. When $g$ is some matrix Lie algebra the conservation laws are just traces of powers of $L$ 

\begin{eqnarray}
\label{eq:e38}
I^{(k)} = \frac{ 1 }{2 } Tr(L^{k}) 
\end{eqnarray}
since trace is invariant under coadjoint action 
\begin{eqnarray}
\frac{ d }{dt }I^{(k)} = \frac{ 1 }{2 } \frac{ d }{dt } Tr(L^{k}) = \frac{ 1 }{2 } Tr(\frac{ d }{dt }L^{k}) = \frac{ k }{2 } Tr(L^{k - 1}\frac{ d }{dt }L) =\nonumber \\\frac{ k }{2 } Tr(L^{k - 1}[L , P]) = \frac{ 1 }{2 } Tr([L^{k}, P]) = 0 
\end{eqnarray}
It is remarkable that each generator of the non-Noether symmetry canonically leads to the Lax pair of a certain type. Such a Lax pairs have definite geometric origin, their Lax matrices are formed by coefficients of invariant tangent valued 1-form on the phase space. In the local coordinates $z_{a}$, where the bivector field $W$, symplectic form $\omega $ and the generator of the symmetry $E$ have the following form 
\begin{eqnarray}
W = \sum _{ab} W_{ab} \frac{ \partial }{\partial z_{a} } \wedge \frac{ \partial }{\partial z_{b} } ~~~~~ \omega = \sum _{ab} \omega _{ab} dz_{a} \wedge dz_{b} ~~~~~ E = \sum _{a} E_{a} \frac{ \partial }{\partial z_{a} } 
\end{eqnarray}
corresponding Lax pair can be calculated explicitly. Namely we have the following theorem: \\
{\bf Theorem 3.} Let $(M , h)$ be regular Hamiltonian system on the $2n$-dimensional Poisson manifold $M$. Then, if the vector field $E$ on $M$ generates the non-Noether symmetry, the following $2n\times 2n$ matrix valued functions on $M$ 

\begin{eqnarray}
\label{eq:e39}
L_{ab} = \sum _{dc} \omega _{ad} \left [ E_{c} \frac{ \partial W_{db} }{\partial z_{c} } - W_{bc} \frac{ \partial E_{d} }{\partial z_{c} } + W_{dc} \frac{ \partial E_{b} }{\partial z_{c} } \right ]\nonumber \\P_{ab} = \sum _{c} \left [ \frac{ \partial W_{bc} }{\partial z_{a} } \frac{ \partial h }{\partial z_{c} } + W_{bc} \frac{ \partial ^{2}h }{\partial z_{a}z_{c} } \right ] 
\end{eqnarray}
form the Lax pair (\ref{eq:e37}) of the dynamical system $(M , h)$. \\
{\bf Proof:} Let us consider the following operator on a space of 1-forms 

\begin{eqnarray}
\label{eq:e40}
\overline{R}_{E}(u) = \Phi _{\omega }([E , \Phi _{W}(u)]) - L_{E}u 
\end{eqnarray}
(here $\Phi _{W}$ and $\Phi _{\omega }$ are maps induced by Poisson bivector field and symplectic form). It is remarkable that $\overline{R}_{E}$ appears to be invariant linear operator. First of all let us show that $\overline{R}_{E}$ is really linear, or in other words, that for arbitrary 1-forms $u$ and $v$ and function $f$$ operator \overline{R}_{E}$ has the following properties 
\begin{eqnarray}
\overline{R}_{E}(u + v) = \overline{R}_{E}(u) + \overline{R}_{E}(v)
\end{eqnarray}
and 
\begin{eqnarray}
\overline{R}_{E}(fu) = f\overline{R}_{E}(u)
\end{eqnarray}
First property is obvious result of linearity of Schouten bracket, Lie derivative and maps $\Phi _{W}$, $\Phi _{\omega }$. Second property can be checked directly 
\begin{eqnarray}
\overline{R}_{E}(fu) = \Phi _{\omega }([E , \Phi _{W}(fu)]) - L_{E}(fu) = \nonumber \\\Phi _{\omega }([E , f\Phi _{W}(u)]) - (L_{E}f)u - fL_{E}u = \nonumber \\\Phi _{\omega }((L_{E}f)\Phi _{W}(u)) + \Phi _{\omega }(f[E , \Phi _{W}(u)]) - (L_{E}f)u - fL_{E}u = \nonumber \\L_{E}f\Phi _{\omega }\Phi _{W}(u) + f\Phi _{\omega }([E , \Phi _{W}(u)]) - (L_{E}f)u - fL_{E}u = \nonumber \\f(\Phi _{\omega }([E , \Phi _{W}(u)]) - L_{E}u) = f\overline{R}_{E}(u) 
\end{eqnarray}
as far as $\Phi _{\omega }\Phi _{W}(u) = u$. Now let us check that $\overline{R}_{E}$ is invariant operator 
\begin{eqnarray}
\frac{ d }{dt }\overline{R}_{E} = L_{X_{h}}\overline{R}_{E} = L_{X_{h}}(\Phi _{\omega }L_{E}\Phi _{W} - L_{E}) = \Phi _{\omega }L_{[X_{h} , E]}\Phi _{W} - L_{[X_{h}, E]} = 0
\end{eqnarray}
because, being Hamiltonian vector field, $X_{h}$ commutes with maps $\Phi _{W}$, $\Phi _{\omega }$ (this is consequence of Liouville theorem) and commutes with $E$ as far as $E$ generates the symmetry $[X_{h}, E] = 0$. In the terms of the local coordinates $\overline{R}_{E}$ has the following form 
\begin{eqnarray}
\overline{R}_{E} = \sum _{ab} L_{ab} dz_{a} \otimes \frac{ \partial }{\partial z_{b} } 
\end{eqnarray}
and the invariance condition 
\begin{eqnarray}
\frac{ d }{dt }\overline{R}_{E} = L_{W(h)}\overline{R}_{E} = 0 
\end{eqnarray}
yields 
\begin{eqnarray}
\frac{ d }{dt }\overline{R}_{E} = \frac{ d }{dt }\sum _{ab} L_{ab} dz_{a} \otimes \frac{ \partial }{\partial z_{b} } = \sum _{ab} \left [ \frac{ d }{dt }L_{ab} \right ] dz_{a} \otimes \frac{ \partial }{\partial z_{b} }\nonumber \\+ \sum _{ab} L_{ab} (L_{W(h)}dz_{a}) \otimes \frac{ \partial }{\partial z_{b} } + \sum _{ab} L_{ab} dz_{a} \otimes \left [ L_{W(h)} \frac{ \partial }{\partial z_{b} } \right ] =\nonumber \\\sum _{ab} \left [\frac{ d }{dt }L_{ab} \right ] dz_{a} \otimes \frac{ \partial }{\partial z_{b} } + \sum _{abcd} L_{ab} \frac{ \partial W_{ad} }{\partial z_{c} } \frac{ \partial h }{\partial z_{d} } dz_{c} \otimes \frac{ \partial }{\partial z_{b} } + \nonumber \\\sum _{abcd} L_{ab} W_{ad} \frac{ \partial ^{2}h }{\partial z_{c}\partial z_{d} } dz_{c} \otimes \frac{ \partial }{\partial z_{b} } + \nonumber \\\sum _{abcd} L_{ab} \frac{ \partial W_{cd} }{\partial z_{b} } \frac{ \partial h }{\partial z_{d} } dz_{a} \otimes \frac{ \partial }{\partial z_{c} } + \sum _{abcd} L_{ab} W_{cd} \frac{ \partial ^{2}h }{\partial z_{b}\partial z_{d} } dz_{a} \otimes \frac{ \partial }{\partial z_{c} } =\nonumber \\\sum _{ab} \left [\frac{ d }{dt }L_{ab} + \sum _{c} (P_{ac}L_{cb} - L_{ac}P_{cb})\right ] dz_{a} \otimes \frac{ \partial }{\partial z_{b} } = 0 
\end{eqnarray}
or in matrix notations 
\begin{eqnarray}
\frac{ d }{dt }L = [L , P]. 
\end{eqnarray}
So, we have proved that the non-Noether symmetry canonically yields a Lax pair on the algebra of linear operators on cotangent bundle over the phase space. \\
{\bf Remark 4.} The conservation laws (\ref{eq:e38}) associated with the Lax pair (\ref{eq:e39}) can be expressed in terms of the integrals of motion $c_{i}$ in quite simple way: 

\begin{eqnarray}
\label{eq:e41}
I^{(k)} = \frac{ 1 }{2 } Tr(L^{k}) = \sum _{i} c_{i}^{k}
\end{eqnarray}
This correspondence follows from the equation (\ref{eq:e18}) and the definition of the operator $\overline{R}_{E}$ (\ref{eq:e40}). One can also write down recursion relation that determines conservation laws $I^{(k)}$ in terms of conservation laws $C^{(k)}$ 

\begin{eqnarray}
\label{eq:e42}
I^{(m)} + (- 1)^{m}mC^{(m)} + \sum ^{ m - 1 }_{k = 1 }(- 1)^{k} I^{(m - k)}C^{(k)} = 0 
\end{eqnarray}
\\
{\bf Sample.} Let us calculate Lax matrix of two particle Toda chain associated with non-Noether symmetry (\ref{eq:e27}). Using (\ref{eq:e39}) it is easy to check that Lax matrix has eight nonzero elements 

\begin{eqnarray}
\label{eq:e43}
L = 
\left( \begin{array}{cccccccccc} z_{1} &0 &0 &- e^{z_{3} - z_{4}} \\ 0 &z_{2} &e^{z_{3} - z_{4}} &0 \\ 0 &1 &z_{1} &0 \\ - 1 &0 &0 &z_{2} \end{array}\right)
\end{eqnarray}
while matrix $P$ involved in Lax pair 
\begin{eqnarray}
\frac{ d }{dt }L = [L , P] 
\end{eqnarray}
has the following form 

\begin{eqnarray}
\label{eq:e44}
P = 
\left( \begin{array}{cccccccccc} 0 &0 &1 &0 \\ 0 &0 &0 &1 \\ - e^{z_{3} - z_{4}} &e^{z_{3} - z_{4}} &0 &0 \\ e^{z_{3} - z_{4}} &- e^{z_{3} - z_{4}} &0 &0 \end{array}\right)
\end{eqnarray}
The conservation laws associated with this Lax pair are total momentum and energy of two particle Toda chain 

\begin{eqnarray}
\label{eq:e45}
I^{(1)} = \frac{ 1 }{2 } Tr(L) = z_{1} + z_{2}\nonumber \\I^{(2)} = \frac{ 1 }{2 } Tr(L^{2}) = z_{1}^{2} + z_{2}^{2} + 2e^{z_{3} - z_{4}} 
\end{eqnarray}
Similarly one can construct Lax matrix of three particle Toda chain, it has 16 nonzero elements 

\begin{eqnarray}
\label{eq:e46}
L = 
\left( \begin{array}{cccccccccc} z_{1} &0 &0 &0 &- e^{z_{4} - z_{5}} &0 \\ 0 &z_{2} &0 &e^{z_{4} - z_{5}} &0 &- e^{z_{5} - z_{6}} \\ 0 &0 &z_{3} &0 &e^{z_{5} - z_{6}} &0 \\ 0 &- 1 &- 1 &z_{1} &0 &0 \\ 1 &0 &- 1 &0 &z_{2} &0 \\ 1 &1 &0 &0 &0 &z_{3} \end{array}\right)
\end{eqnarray}
with matrix $P$ 

\begin{eqnarray}
\label{eq:e47}
P = 
\left( \begin{array}{cccccccccc} 0 &0 &0 &1 &0 &0 \\ 0 &0 &0 &0 &1 &0 \\ 0 &0 &0 &0 &0 &1 \\ - e^{z_{4} - z_{5}} &e^{z_{4} - z_{5}} &0 &0 &0 &0 \\ e^{z_{4} - z_{5}} &- e^{z_{4} - z_{5}} - e^{z_{5} - z_{6}} &e^{z_{5} - z_{6}} &0 &0 &0 \\ 0 &e^{z_{5} - z_{6}} &- e^{z_{5} - z_{6}} &0 &0 &0 \end{array}\right)
\end{eqnarray}
Corresponding conservation laws reproduce total momentum, energy and second Hamiltonian involved in bi-Hamiltonian realization of Toda chain 

\begin{eqnarray}
\label{eq:e48}
I^{(1)} = \frac{ 1 }{2 } Tr(L) = z_{1} + z_{2}\nonumber \\I^{(2)} = \frac{ 1 }{2 } Tr(L^{2}) = z_{1}^{2} + z_{2}^{2} + z_{3}^{2} + 2e^{z_{4} - z_{5}} + 2e^{z_{5} - z_{6}}\nonumber \\I^{(3)} = \frac{ 1 }{2 } Tr(L^{3}) = z_{1}^{3} + z_{2}^{3} + z_{3}^{3} + 3(z_{1} + z_{2})e^{z_{4} - z_{5}} + 3(z_{2} + z_{3})e^{z_{5} - z_{6}} 
\end{eqnarray}
\\
\section{Involutivity of conservation laws}
Now let us focus on the integrability issues. We know that $n$ integrals of motion are associated with each generator of non-Noether symmetry, in the same time we know that, according to the Liouville-Arnold theorem, regular Hamiltonian system $(M, h)$ on $2n$ dimensional symplectic manifold $M$ is completely integrable (can be solved completely) if it admits $n$ functionally independent integrals of motion in involution. One can understand functional independence of set of conservation laws $c_{1}, c_{2} ... c_{n}$ as linear independence of either differentials of conservation laws $dc_{1}, dc_{2} ... dc_{n}$ or corresponding Hamiltonian vector fields $X_{c_{1}}, X_{c_{2}} ... X_{c_{n}}$. Strictly speaking we can say that conservation laws $c_{1}, c_{2} ... c_{n}$ are functionally independent if Lesbegue measure of the set of points of phase space $M$ where differentials $dc_{1}, dc_{2} ... dc_{n}$ become linearly dependent is zero. Involutivity of conservation laws means that all possible Poisson brackets of these conservation laws vanish pair wise 
\begin{eqnarray}
\{c_{i} , c_{j}\} = 0 ~~~~~ i, j = 1... n
\end{eqnarray}
In terms of the vector fields, existence of involutive family of $n$ functionally independent conservation laws $c_{1}, c_{2} ... c_{n}$ implies that corresponding Hamiltonian vector fields $X_{c_{1}}, X_{c_{2}} ... X_{c_{n}}$ span Lagrangian subspace (isotropic subspace of dimension $n$) of tangent space (at each point of $M$). Indeed, due to property (\ref{eq:e10}) 
\begin{eqnarray}
\{c_{i} , c_{j}\} = \omega (X_{c_{i}} , X_{c_{j}}) = 0
\end{eqnarray}
thus space spanned by $X_{c_{1}}, X_{c_{2}} ... X_{c_{n}}$ is isotropic. Dimension of this space is $n$ so it is Lagrangian. Note also that distribution $X_{c_{1}}, X_{c_{2}} ... X_{c_{n}}$ is integrable since due to (\ref{eq:e9}) 
\begin{eqnarray}
[X_{c_{i}} , X_{c_{j}}] = X_{\{c_{i} , c_{j}\}} = 0
\end{eqnarray}
and according to Frobenius theorem there exists submanifold of $M$ such that distribution $X_{c_{1}}, X_{c_{2}} ... X_{c_{n}}$ spans tangent space of this submanifold. Thus for phase space geometry existence of complete involutive set of integrals of motion implies existence of invariant Lagrangian submanifold.\\
Now let us look at conservation laws $Y^{(1)}, Y^{(2)} ... Y^{(n)}$ associated with generator of non-Noether symmetry. Generally speaking these conservation laws might appear to be neither functionally independent nor involutive. However it is reasonable to ask the question -- what condition should be satisfied by the generator of the non-Noether symmetry to ensure the involutivity ($\{Y^{(k)} , Y^{(m)}\} = 0$) of conserved quantities? In Lax theory situation is very similar --- each Lax matrix leads to the set of conservation laws but in general this set is not involutive, however in Lax theory there is certain condition known as Classical Yang-Baxter Equation (CYBE) that being satisfied by Lax matrix ensures that conservation laws are in involution. Since involutivity of the conservation laws is closely related to the integrability, it is essential to have some analog of CYBE for the generator of non-Noether symmetry. To address this issue we would like to propose the following theorem. \\
{\bf Theorem 4.} If the vector field $E$ on $2n$-dimensional Poisson manifold $M$ satisfies the condition 

\begin{eqnarray}
\label{eq:e49}
[[E[E , W]]W] = 0
\end{eqnarray}
and $W$ bivector field has maximal rank ($W^{n} \neq 0$) then the functions (\ref{eq:e16}) are in involution 
\begin{eqnarray}
\{Y^{(k)} , Y^{(m)}\} = 0
\end{eqnarray}
\\
{\bf Proof:} First of all let us note that the identity (\ref{eq:e5}) satisfied by the Poisson bivector field $W$ is responsible for the Liouville theorem 

\begin{eqnarray}
\label{eq:e50}
[W , W] = 0 ~~~~~ \Leftrightarrow ~~~~~ L_{W(f)}W = [W(f) , W] = 0
\end{eqnarray}
that follows from the graded Jacoby identity satisfied by Schouten bracket. By taking the Lie derivative of the expression (\ref{eq:e5}) we obtain another useful identity 
\begin{eqnarray}
L_{E}[W , W] = [E[W , W]] = [[E , W] W] + [W[E , W]] = 2[\hat{W} , W] = 0.
\end{eqnarray}
This identity gives rise to the following relation 

\begin{eqnarray}
\label{eq:e51}
[\hat{W} , W] = 0 ~~~~~\Leftrightarrow ~~~~~ [\hat{W}(f) , W] = - [\hat{W} , W(f)]
\end{eqnarray}
and finally condition (\ref{eq:e49}) ensures third identity 
\begin{eqnarray}
[\hat{W} , \hat{W}] = 0
\end{eqnarray}
yielding Liouville theorem for $\hat{W}$ 

\begin{eqnarray}
\label{eq:e52}
[\hat{W} , \hat{W}] = 0 ~~~~~\Leftrightarrow ~~~~~ [\hat{W}(f) , \hat{W}] = 0
\end{eqnarray}
Indeed 
\begin{eqnarray}
[\hat{W} , \hat{W}] = [[E , W]\hat{W}] = [[\hat{W} , E]W] = \nonumber \\
- [[E , \hat{W}]W] = - [[E[E , W]]W] = 0
\end{eqnarray}
Now let us consider two different solutions $c_{i} \neq c_{j}$ of the equation (\ref{eq:e18}). By taking the Lie derivative of the equation 
\begin{eqnarray}
(\hat{W} - c_{i}W)^{n} = 0
\end{eqnarray}
along the vector fields $W(c_{j})$ and $\hat{W}(c_{j})$ and using Liouville theorem for $W$ and $\hat{W}$ bivectors we obtain the following relations 

\begin{eqnarray}
\label{eq:e53}
(\hat{W} - c_{i}W)^{n - 1}(L_{W(c_{j})}\hat{W} - \{c_{j} , c_{i}\}W) = 0,
\end{eqnarray}
and 

\begin{eqnarray}
\label{eq:e54}
(\hat{W} - c_{i}W)^{n - 1}(c_{i}L_{\hat{W}(c_{j})}W + \{c_{j} , c_{i}\}_{\bullet }W) = 0,
\end{eqnarray}
where 
\begin{eqnarray}
\{c_{i} , c_{j}\}_{\bullet } = \hat{W}(dc_{i} \wedge dc_{j})
\end{eqnarray}
is the Poisson bracket calculated by means of the bivector field $\hat{W}$. Now multiplying (\ref{eq:e53}) by $c_{i}$ subtracting (\ref{eq:e54}) and using identity (\ref{eq:e51}) gives rise to 

\begin{eqnarray}
\label{eq:e55}
(\{c_{i} , c_{j}\}_{\bullet } - c_{i}\{c_{i} , c_{j}\})(\hat{W} - c_{i}W)^{n - 1}W = 0
\end{eqnarray}
Thus, either 

\begin{eqnarray}
\label{eq:e56}
\{c_{i} , c_{j}\}_{\bullet } - c_{i}\{c_{i} , c_{j}\} = 0
\end{eqnarray}
or the volume field $(\hat{W} - c_{i}W)^{n - 1}W$ vanishes. In the second case we can repeat (\ref{eq:e53})-(\ref{eq:e55}) procedure for the volume field $(\hat{W} - c_{i}W)^{n - 1}W$ yielding after $n$ iterations $W^{n} = 0$ that according to our assumption (that the dynamical system is regular) is not true. As a result we arrived at (\ref{eq:e56}) and by the simple interchange of indices $i \leftrightarrow j$ we get 

\begin{eqnarray}
\label{eq:e57}
\{c_{i} , c_{j}\}_{\bullet } - c_{j}\{c_{i} , c_{j}\} = 0
\end{eqnarray}
Finally by comparing (\ref{eq:e56}) and (\ref{eq:e57}) we obtain that the functions $c_{i}$ are in involution with respect to the both Poisson structures (since $c_{i} \neq c_{j}$) 
\begin{eqnarray}
\{c_{i} , c_{j}\}_{\bullet } = \{c_{i} , c_{j}\} = 0
\end{eqnarray}
and according to (\ref{eq:e19}) the same is true for the integrals of motion $Y^{(k)}$. \\
{\bf Remark 5.} Theorem 4 is useful in multidimensional dynamical systems where involutivity of conservation laws can not be checked directly.\\
\section{Bi-Hamiltonian systems}
Further we will focus on non-Noether symmetries that satisfy condition (\ref{eq:e49}). Besides yielding involutive families of conservation laws, such a symmetries appear to be related to many known geometric structures such as bi-Hamiltonian systems and Fr\"{o}licher-Nijenhuis operators (torsionless tangent valued differential 1-forms). The relationship between non-Noether symmetries and bi-Hamiltonian structures was already implicitly outlined in the proof of Theorem 4. Now let us pay more attention to this issue. \\
Originally bi-Hamiltonian structures were introduced by F. Magri in analisys of integrable infinite dimensional Hamiltonian systems such as Korteweg-de Vries (KdV) and modified Korteweg-de Vries (mKdV) hierarchies, Nonlinear Schr\"{o}dinger equation and Harry Dym equation. Since that time bi-Hahmiltonian formalism is effectively used in construction of involutive families of conservation laws in integrable models\\
Generic bi-Hamiltonian structure on $2n$ dimensional manifold consists out of two Poisson bivector fields $W$ and $\hat{W}$ satisfying certain compatibility condition $[\hat{W} , W] = 0$. If, in addition, one of these bivector fields is nondegenerate ($W^{n} \neq 0$) then bi-Hamiltonian system is called regular. Further we will discuss only regular bi-Hamiltonian systems. Note that each Poisson bivector field by definition satisfies condition (\ref{eq:e5}). So we actually impose four restrictions on bivector fields $W$ and $\hat{W}$ 

\begin{eqnarray}
\label{eq:e58}
[W , W] = [\hat{W} , W] = [\hat{W} , \hat{W}] = 0
\end{eqnarray}
and 

\begin{eqnarray}
\label{eq:e59}
W^{n} \neq 0
\end{eqnarray}
During the proof of Theorem 4 we already showed that bivector fields $W$ and $\hat{W} = [E , W]$ satisfy conditions (\ref{eq:e58}) (see (\ref{eq:e50})-(\ref{eq:e52})), thus we can formulate the following statement \\
{\bf Theorem 5.} Let $(M , h)$ be regular Hamiltonian system on the $2n$-dimensional manifold $M$ endowed with regular Poisson bivector field $W$. Then, if the vector field $E$ on $M$ generates the non-Noether symmetry, and satisfies condition 
\begin{eqnarray}
[[E[E , W]]W] = 0,
\end{eqnarray}
the following bivector fields on $M$ 
\begin{eqnarray}
W, \hat{W} = [E , W]
\end{eqnarray}
form invariant bi-Hamiltonian system ($[W , W] = [\hat{W} , W] = [\hat{W} , \hat{W}] = 0$). \\
{\bf Proof:} See proof of Theorem 4.\\
{\bf Sample.} One can check that the non-Noether symmetry (\ref{eq:e27}) satisfies condition (\ref{eq:e49}) while bivector fields 
\begin{eqnarray}
W = \frac{ \partial }{\partial z_{1} } \wedge \frac{ \partial }{\partial z_{3} } + \frac{ \partial }{\partial z_{2} } \wedge \frac{ \partial }{\partial z_{4} }
\end{eqnarray}
and 
\begin{eqnarray}
\hat{W} = [E , W] = z_{1} \frac{ \partial }{\partial z_{1} } \wedge \frac{ \partial }{\partial z_{3} } + z_{2} \frac{ \partial }{\partial z_{2} } \wedge \frac{ \partial }{\partial z_{4} } +\nonumber \\
e^{z_{3} - z_{4}} \frac{ \partial }{\partial z_{1} } \wedge \frac{ \partial }{\partial z_{2} } + \frac{ \partial }{\partial z_{3} } \wedge \frac{ \partial }{\partial z_{4} } 
\end{eqnarray}
form bi-Hamiltonian system $[W , W] = [W , \hat{W}] = [\hat{W} , \hat{W}] = 0$. Similarly, one can recover bi-Hamiltonian system of three particle Toda chain associated with symmetry (\ref{eq:e32}). It is formed by bivector fields 
\begin{eqnarray}
W = \frac{ \partial }{\partial z_{1} } \wedge \frac{ \partial }{\partial z_{4} } + \frac{ \partial }{\partial z_{2} } \wedge \frac{ \partial }{\partial z_{5} } + \frac{ \partial }{\partial z_{3} } \wedge \frac{ \partial }{\partial z_{6} } 
\end{eqnarray}
and 
\begin{eqnarray}
\hat{W} = [E , W] = z_{1} \frac{ \partial }{\partial z_{1} } \wedge \frac{ \partial }{\partial z_{4} } + z_{2} \frac{ \partial }{\partial z_{2} } \wedge \frac{ \partial }{\partial z_{5} } + z_{3} \frac{ \partial }{\partial z_{3} } \wedge \frac{ \partial }{\partial z_{6} } +\nonumber \\e^{z_{4} - z_{5}} \frac{ \partial }{\partial z_{1} } \wedge \frac{ \partial }{\partial z_{2} } + e^{z_{5} - z_{6}} \frac{ \partial }{\partial z_{2} } \wedge \frac{ \partial }{\partial z_{3} } + \nonumber \\\frac{ \partial }{\partial z_{3} } \wedge \frac{ \partial }{\partial z_{4} } + \frac{ \partial }{\partial z_{4} } \wedge \frac{ \partial }{\partial z_{5} } + \frac{ \partial }{\partial z_{5} } \wedge \frac{ \partial }{\partial z_{6} } 
\end{eqnarray}
\\
In terms of differential forms bi-Hamiltonian structure is formed by couple of closed differential 2-forms: symplectic form $\omega $ (such that $d\omega = 0$ and $\omega ^{n} \neq 0$) and $\omega ^{\bullet } = L_{E}\omega $ (clearly $d\omega ^{\bullet } = dL_{E}\omega = L_{E}d\omega = 0$). It is important that by taking Lie derivative of Hamilton's equation 
\begin{eqnarray}
i_{X_{h}}\omega + dh = 0
\end{eqnarray}
along the generator $E$ of symmetry 
\begin{eqnarray}
L_{E}(i_{X_{h}}\omega + dh) = i_{[E , X_{h}]}\omega + i_{X_{h}}L_{E}\omega + L_{E}dh = i_{X_{h}}\omega ^{\bullet } + dL_{E}h = 0
\end{eqnarray}
one obtains another Hamilton's equation 
\begin{eqnarray}
i_{X_{h}}\omega ^{\bullet } + dh^{\bullet } = 0
\end{eqnarray}
where $h^{\bullet } = L_{E}h$. This is actually second Hamiltonian realization of equations of motion and thus under certain conditions existence of non-Noether symmetry gives rise to additional presymplectic structure $\omega ^{\bullet }$ and additional Hamiltonian realization of the dynamical system. In many integrable models admitting bi-Hamiltonian realization (including Toda chain, Korteweg-de Vries hierarchy, Nonlinear Schr\"{o}dinger equation, Broer-Kaup system and Benney system) non-Noether symmetries that are responsible for existence of bi-Hamiltonian structures has been found and motivated further investigation of relationship between symmetries and bi-Hamiltonian structures. Namely it seems to be interesting to know whether in general case existence of bi-Hamiltonian structure is related to non-Noether symmetry. Let us consider more general case and suppose that we have couple of differential 2-forms $\omega $ and $\omega ^{\bullet }$ such that 
\begin{eqnarray}
d\omega = d\omega ^{\bullet } = 0, ~~~~~ \omega ^{n} \neq 0
\end{eqnarray}
\begin{eqnarray}
i_{X_{h}}\omega + dh = 0
\end{eqnarray}
and 
\begin{eqnarray}
i_{X_{h}}\omega ^{\bullet } + dh^{\bullet } = 0
\end{eqnarray}
The question is whether there exists vector field $E$ (generator of non-Noether symmetry) such that $[E , X_{h}] = 0$ and $\omega ^{\bullet } = L_{E}\omega $. \\
The answer depends on $\omega ^{\bullet }$. Namely if $\omega ^{\bullet }$ is exact form (there exists 1-form $\theta ^{\bullet }$ such that $\omega ^{\bullet } = d\theta ^{\bullet }$) then one can argue that such a vector field exists and thus any exact bi-Hamiltonian structure is related to hidden non-Noether symmetry. To outline proof of this statement let us introduce vector field $E^{\bullet }$ defined by 
\begin{eqnarray}
i_{E^{\bullet }}\omega = \theta ^{\bullet }
\end{eqnarray}
(such a vector field always exist because $\omega $ is nondegenerate 2-form). By construction 
\begin{eqnarray}
L_{E^{\bullet }} \omega = \omega ^{\bullet }
\end{eqnarray}
Indeed 
\begin{eqnarray}
L_{E^{\bullet }}\omega = di_{E^{\bullet }}\omega + i_{E^{\bullet }}d\omega = d\theta ^{\bullet } = \omega ^{\bullet }
\end{eqnarray}
And 
\begin{eqnarray}
i_{[E^{\bullet }, X_{h}]}\omega = L_{E^{\bullet }}(i_{X_{h}}\omega ) - i_{X_{h}}L_{E^{\bullet }}\omega = - d(E^{\bullet }(h) - h^{\bullet }) = - dh'
\end{eqnarray}
In other words $[X_{h} , E^{\bullet }]$ is Hamiltonian vector field 
\begin{eqnarray}
[X_{h} , E] = X_{h'}
\end{eqnarray}
One can also construct locally Hamiltonian vector field $X_{g}$, that satisfies the same commutation relation. Namely let us define function (in general case it can be done only locally) 
\begin{eqnarray}
g(z) = \int ^{ t }_{0 } h'dt
\end{eqnarray}
where integration along solution of Hamilton's equation, with fixed origin and end point in $z(t) = z$, is assumed. And then it is easy to verify that locally Hamiltonian vector field associated with $g(z)$, by construction, satisfies the same commutation relations as $E^{\bullet }$ (namely $[X_{h} , X_{g}] = X_{h'}$). Using $E^{\bullet }$ and $X_{h'}$ one can construct generator of non-Noether symmetry --- non-Hamiltonian vector field $E = E^{\bullet } - X_{g}$ commuting with $X_{h}$ and satisfying 
\begin{eqnarray}
L_{E}\omega = L_{E^{\bullet }}\omega - L_{X_{g}}\omega = L_{E^{\bullet }}\omega = \omega ^{\bullet }
\end{eqnarray}
(thanks to Liouville's theorem $L_{X_{g}}\omega = 0$). So in case of regular Hamiltonian system every exact bi-Hamiltonian structure is naturally associated with some (non-Noether) symmetry of space of solutions. In case when bi-Hamiltonian structure is not exact ($\omega ^{\bullet }$ is closed but not exact) then due to 
\begin{eqnarray}
\omega ^{\bullet } = L_{E}\omega = di_{E}\omega + i_{E}d\omega = di_{E}\omega 
\end{eqnarray}
it is clear that such a bi-Hamiltonian system is not related to symmetry. However in all known cases bi-Hamiltonian structures seem to be exact. \\
\section{Bidifferential calculi}
Another important concept that is often used in theory of dynamical systems and may be related to the non-Noether symmetry is the bidifferential calculus (bicomplex approach). Recently A.~Dimakis and F.~M\"{u}ller-Hoissen applied bidifferential calculi to the wide range of integrable models including KdV hierarchy, KP equation, self-dual Yang-Mills equation, Sine-Gordon equation, Toda models, non-linear Schr\"{o}dinger and Liouville equations. It turns out that these models can be effectively described and analyzed using the bidifferential calculi \cite{r17} \cite{r24}. Here we would like to show that each generator of non-Noether symmetry satisfying condition $[[E[E , W]]W] = 0$ gives rise to certain bidifferential calculus. \\
Before we proceed let us specify what kind of bidifferential calculi we plan to consider. Under the bidifferential calculus we mean the graded algebra of differential forms over the phase space 

\begin{eqnarray}
\label{eq:e60}
\Omega = \cup ^{ \infty }_{k = 0 } \Omega ^{(k)} 
\end{eqnarray}
($\Omega ^{(k)}$ denotes the space of $k$-degree differential forms) equipped with a couple of differential operators 

\begin{eqnarray}
\label{eq:e61}
d, \bar{d} : \Omega ^{(k)} \rightarrow \Omega ^{(k + 1)} 
\end{eqnarray}
satisfying $d^{2} = \bar{d}^{2} = d\bar{d} + \bar{d}d = 0$ conditions (see \cite{r24}). In other words we have two De Rham complexes $M, \Omega , d$ and $M, \Omega , \bar{d}$ on algebra of differential forms over the phase space. And these complexes satisfy certain compatibility condition --- their differentials anticommute with each other $d\bar{d} + \bar{d}d = 0$. Now let us focus on non-Noether symmetries. It is interesting that if generator of the non-Noether symmetry satisfies equation $[[E[E , W]]W] = 0$ then we are able to construct an invariant bidifferential calculus of a certain type. This construction is summarized in the following theorem: \\
{\bf Theorem 6.} Let $(M , h)$ be regular Hamiltonian system on the Poisson manifold $M$. Then, if the vector field $E$ on $M$ generates the non-Noether symmetry and satisfies the equation 
\begin{eqnarray}
[[E[E , W]]W] = 0,
\end{eqnarray}
the differential operators 

\begin{eqnarray}
\label{eq:e62}
du = \Phi _{\omega }([W , \Phi _{W}(u)])
\end{eqnarray}

\begin{eqnarray}
\label{eq:e63}
\bar{d}u = \Phi _{\omega }([[E , W]\Phi _{W}(u)])
\end{eqnarray}
form invariant bidifferential calculus ($d^{2} = \bar{d}^{2} = d\bar{d} + \bar{d}d = 0$) over the graded algebra of differential forms on $M$. \\
{\bf Proof:} First of all we have to show that $d$ and $\bar{d}$ are really differential operators , i.e., they are linear maps from $\Omega ^{(k)}$ into $\Omega ^{(k + 1)}$, satisfy derivation property and are nilpotent ($d^{2} = \bar{d}^{2} = 0$). Linearity is obvious and follows from the linearity of the Schouten bracket $[ , ]$ and $\Phi _{W}, \Phi _{\omega }$ maps. Then, if $u$ is a $k$-degree form $\Phi _{W}$ maps it on $k$-degree multivector field and the Schouten brackets $[W , \Phi _{W}(u)]$ and $[[E , W]\Phi _{W}(u)]$ result the $k + 1$-degree multivector fields that are mapped on $k + 1$-degree differential forms by $\Phi _{\omega }$. So, $d$ and $\bar{d}$ are linear maps from $\Omega ^{(k)}$ into $\Omega ^{(k + 1)}$. Derivation property follows from the same feature of the Schouten bracket $[ , ]$ and linearity of $\Phi _{W}$ and $\Phi _{\omega }$ maps. Now we have to prove the nilpotency of $d$ and $\bar{d}$. Let us consider $d^{2}u$ 
\begin{eqnarray}
d^{2}u = \Phi _{\omega }([W , \Phi _{W}(\Phi _{\omega }([W , \Phi _{W}(u)]))]) = \Phi _{\omega }([W[W , \Phi _{W}(u)]]) = 0
\end{eqnarray}
as a result of the property (\ref{eq:e50}) and the Jacoby identity for $[ , ]$ bracket. In the same manner 
\begin{eqnarray}
\bar{d}^{2}u = \Phi _{\omega }([[W , E][[W , E]\Phi _{W}(u)]]) = 0
\end{eqnarray}
according to the property (\ref{eq:e52}) of $[W , E] = \hat{W}$ and the Jacoby identity. Thus, we have proved that $d$ and $\bar{d}$ are differential operators (in fact $d$ is ordinary exterior differential and the expression (\ref{eq:e62}) is its well known representation in terms of Poisson bivector field). It remains to show that the compatibility condition $d\bar{d} + \bar{d}d = 0$ is fulfilled. Using definitions of $d, \bar{d}$ and the Jacoby identity we get 
\begin{eqnarray}
(d\bar{d} + \bar{d}d)(u) = \Phi _{\omega }([[[W , E]W]\Phi _{W}(u)]) = 0 
\end{eqnarray}
as far as (\ref{eq:e51}) is satisfied. So, $d$ and $\bar{d}$ form the bidifferential calculus over the graded algebra of differential forms. It is also clear that the bidifferential calculus $d, \bar{d}$ is invariant, since both $d$ and $\bar{d}$ commute with time evolution operator $W(h) = \{h, \}$. \\
{\bf Remark 6.} Conservation laws that are associated with the bidifferential calculus (\ref{eq:e62}) (\ref{eq:e63}) and form Lenard scheme (see \cite{r24}): 
\begin{eqnarray}
(k + 1)\bar{d}I^{(k)} = kdI^{(k + 1)}
\end{eqnarray}
coincide with the sequence of integrals of motion (\ref{eq:e41}). Proof of this correspondence lays outside the scope of present manuscript, but can be done in the manner similar to \cite{r17}. \\
{\bf Sample.} The symmetry (\ref{eq:e27}) endows $R^{4}$ with bicomplex structure $d, \bar{d}$ where $d$ is ordinary exterior derivative while $\bar{d}$ is defined by 

\begin{eqnarray}
\label{eq:e64}
\bar{d}z_{1} = z_{1}dz_{1} - e^{z_{3} - z_{4}}dz_{4}\nonumber \\\bar{d}z_{2} = z_{2}dz_{2} + e^{z_{3} - z_{4}}dz_{3}\nonumber \\\bar{d}z_{3} = z_{1}dz_{3} + dz_{2}\nonumber \\\bar{d}z_{4} = z_{2}dz_{4} - dz_{1} 
\end{eqnarray}
and is extended to whole De Rham complex by linearity, derivation property and compatibility property $d\bar{d} + \bar{d}d = 0$. By direct calculations one can verify that calculus constructed in this way is consistent and satisfies $\bar{d}^{2} = 0$ property. To illustrate technique let us explicitly check that $\bar{d}^{2}z_{1} = 0$. Indeed 
\begin{eqnarray}
\bar{d}^{2}z_{1} = \bar{d}\bar{d}z_{1} = \bar{d}(z_{1}dz_{1} - e^{z_{3} - z_{4}}dz_{4}) = \bar{d}z_{1} \wedge dz_{1} + z_{1}\bar{d}dz_{1} \nonumber \\- e^{z_{3} - z_{4}}\bar{d}z_{3} \wedge dz_{4} + e^{z_{3} - z_{4}}\bar{d}z_{4} \wedge dz_{4} - e^{z_{3} - z_{4}}\bar{d}dz_{4} = \nonumber \\\bar{d}z_{1} \wedge dz_{1} - z_{1}d\bar{d}z_{1} - e^{z_{3} - z_{4}}\bar{d}z_{3} \wedge dz_{4} \nonumber \\+ e^{z_{3} - z_{4}}\bar{d}z_{4} \wedge dz_{4} + e^{z_{3} - z_{4}}d\bar{d}z_{4} = 0 
\end{eqnarray}
Because of properties 
\begin{eqnarray}
\bar{d}z_{1} \wedge dz_{1} = e^{z_{3} - z_{4}}dz_{1} \wedge dz_{4},
\end{eqnarray}
\begin{eqnarray}
- z_{1}d\bar{d}z_{1} = z_{1}e^{z_{3} - z_{4}}dz_{3} \wedge dz_{4},
\end{eqnarray}
\begin{eqnarray}
- e^{z_{3} - z_{4}}\bar{d}z_{3} \wedge dz_{4} = - z_{1}e^{z_{3} - z_{4}}dz_{1} \wedge dz_{4} - e^{z_{3} - z_{4}}dz_{2} \wedge dz_{4},
\end{eqnarray}
\begin{eqnarray}
e^{z_{3} - z_{4}}\bar{d}z_{4} \wedge dz_{4} = e^{z_{3} - z_{4}}dz_{2} \wedge dz_{4}
\end{eqnarray}
and 
\begin{eqnarray}
e^{z_{3} - z_{4}}d\bar{d}z_{4} = - e^{z_{3} - z_{4}}dz_{1} \wedge dz_{4} 
\end{eqnarray}
Similarly one can show that 
\begin{eqnarray}
\bar{d}^{2}z_{2} = \bar{d}^{2}z_{3} = \bar{d}^{2}z_{4} = 0
\end{eqnarray}
and thus $\bar{d}$ is nilpotent operator $\bar{d}^{2} = 0$. Note also that conservation laws 
\begin{eqnarray}
I^{(1)} = z_{1} + z_{2}\nonumber \\I^{(2)} = z_{1}^{2} + z_{2}^{2} + 2e^{z_{3} - z_{4}} 
\end{eqnarray}
form the simplest Lenard scheme 
\begin{eqnarray}
2\bar{d}I^{(1)} = dI^{(2)}
\end{eqnarray}
Similarly one can construct bidifferential calculus associated with non-Noether symmetry (\ref{eq:e32}) of three particle Toda chain. In this case $\bar{d}$ can be defined by 

\begin{eqnarray}
\label{eq:e65}
\bar{d}z_{1} = z_{1}dz_{1} - e^{z_{4} - z_{5}}dz_{5}\nonumber \\\bar{d}z_{2} = z_{2}dz_{2} + e^{z_{4} - z_{5}}dz_{4} - e^{z_{5} - z_{6}}dz_{6}\nonumber \\\bar{d}z_{3} = z_{3}dz_{3} + e^{z_{5} - z_{6}}dz_{5}\nonumber \\\bar{d}z_{4} = z_{1}dz_{4} - dz_{2} - dz_{3}\nonumber \\\bar{d}z_{5} = z_{2}dz_{5} + dz_{1} - dz_{3}\nonumber \\\bar{d}z_{6} = z_{3}dz_{6} + dz_{1} + dz_{2} 
\end{eqnarray}
and as in case of two particle Toda it can be extended to whole De Rham complex by linearity, derivation property and compatibility property $d\bar{d} + \bar{d}d = 0$. One can check that conservation laws of Toda chain 
\begin{eqnarray}
I^{(1)} = z_{1} + z_{2}\nonumber \\I^{(2)} = z_{1}^{2} + z_{2}^{2} + z_{3}^{2} + 2e^{z_{4} - z_{5}} + 2e^{z_{5} - z_{6}}\nonumber \\I^{(3)} = z_{1}^{3} + z_{2}^{3} + z_{3}^{3} + 3(z_{1} + z_{2})e^{z_{4} - z_{5}} + 3(z_{2} + z_{3})e^{z_{5} - z_{6}} 
\end{eqnarray}
form Lenard scheme 
\begin{eqnarray}
2\bar{d}I^{(1)} = dI^{(2)}
\end{eqnarray}
\begin{eqnarray}
3\bar{d}I^{(2)} = 2dI^{(3)}
\end{eqnarray}
\\
\section{Fr\"{o}licher-Nijenhuis geometry}
Finally we would like to reveal some features of the operator $\overline{R}_{E}$ (\ref{eq:e40}) and to show how Fr\"{o}licher-Nijenhuis geometry arises in Hamiltonian system that possesses certain non-Noether symmetry. From the geometric properties of the tangent valued forms we know that the traces of powers of a linear operator $F$ on tangent bundle are in involution whenever its Fr\"{o}licher-Nijenhuis torsion $T(F)$ vanishes, i. e. whenever for arbitrary vector fields $X,Y$ the condition 
\begin{eqnarray}
T(F)(X , Y) = [FX , FY] - F([FX , Y] + [X , FY] - F[X , Y]) = 0
\end{eqnarray}
is satisfied. Torsionless forms are also called Fr\"{o}licher-Nijenhuis operators and are widely used in theory of integrable models, where they play role of recursion operators and are used in construction of involutive family of conservation laws. We would like to show that each generator of non-Noether symmetry satisfying equation $[[E[E , W]]W] = 0$ canonically leads to invariant Fr\"{o}licher-Nijenhuis operator on tangent bundle over the phase space. This operator can be expressed in terms of generator of symmetry and isomorphism defined by Poisson bivector field. Strictly speaking we have the following theorem. \\
{\bf Theorem 7.} Let $(M , h)$ be regular Hamiltonian system on the Poisson manifold $M$. If the vector field $E$ on $M$ generates the non-Noether symmetry and satisfies the equation 
\begin{eqnarray}
[[E[E , W]]W] = 0
\end{eqnarray}
then the linear operator, defined for every vector field $X$ by equation 

\begin{eqnarray}
\label{eq:e66}
R_{E}(X) = \Phi _{W}(L_{E}\Phi _{\omega }(X)) - [E , X] 
\end{eqnarray}
is invariant Fr\"{o}licher-Nijenhuis operator on $M$. \\
{\bf Proof.} Invariance of $R_{E}$ follows from the invariance of the $\overline{R}_{E}$ defined by (\ref{eq:e40}) (note that for arbitrary 1-form vector field $u$ and vector field $X$ contraction $i_{X}u$ has the property $i_{R_{E}X}u = i_{X}\overline{R}_{E}u$, so $R_{E}$ is actually transposed to $\overline{R}_{E}$). It remains to show that the condition (\ref{eq:e49}) ensures vanishing of the Fr\"{o}licher-Nijenhuis torsion $T(R_{E})$ of $R_{E}$, i.e. for arbitrary vector fields $X, Y$ we must get 

\begin{eqnarray}
\label{eq:e67}
T(R_{E})(X , Y) = [R_{E}(X) , R_{E}(Y)] - R_{E}([R_{E}(X) , Y] + [X , R_{E}(Y)] - \nonumber \\
R_{E}([X , Y])) = 0 
\end{eqnarray}
First let us introduce the following auxiliary 2-forms 

\begin{eqnarray}
\label{eq:e68}
\omega = \Phi _{\omega }(W), ~~~~~ \omega ^{\bullet } = \overline{R}_{E}\omega ~~~~~ \omega ^{\bullet \bullet } = \overline{R}_{E}\omega ^{\bullet }
\end{eqnarray}
Using the realization (\ref{eq:e62}) of the differential $d$ and the property (\ref{eq:e5}) yields 
\begin{eqnarray}
d\omega = \Phi _{\omega }([W , W]) = 0
\end{eqnarray}
Similarly, using the property (\ref{eq:e51}) we obtain 
\begin{eqnarray}
d\omega ^{\bullet } = d\Phi _{\omega }([E , W]) - dL_{E}\omega = \Phi _{\omega }([[E , W]W]) - L_{E}d\omega = 0
\end{eqnarray}
And finally, taking into account that $\omega ^{\bullet } = 2\Phi _{\omega }([E , W])$ and using the condition (\ref{eq:e49}), we get 
\begin{eqnarray}
d\omega ^{\bullet \bullet } = 2\Phi _{\omega }([[E[E , W]]W]) - 2dL_{E}\omega ^{\bullet } = - 2L_{E}d\omega ^{\bullet } = 0
\end{eqnarray}
So the differential forms $\omega , \omega ^{\bullet }, \omega ^{\bullet \bullet }$ are closed 

\begin{eqnarray}
\label{eq:e69}
d\omega = d\omega ^{\bullet } = d\omega ^{\bullet \bullet } = 0
\end{eqnarray}
Now let us consider the contraction of $T(R_{E})$ and $\omega $. 

\begin{eqnarray}
\label{eq:e70}
i_{T(R_{E})(X , Y)}\omega = i_{[R_{E}X , R_{E}Y]}\omega - i_{[R_{E}X , Y]}\omega ^{\bullet } - i_{[X , R_{E}Y]}\omega ^{\bullet } + i_{[X , Y]}\omega ^{\bullet \bullet } = \nonumber \\
L_{R_{E}X}i_{Y}\omega ^{\bullet } - i_{R_{E}Y}L_{X}\omega ^{\bullet } - \nonumber \\
L_{R_{E}X}i_{Y}\omega ^{\bullet } + i_{Y}L_{R_{E}X}\omega ^{\bullet } - L_{X}i_{R_{E}Y}\omega ^{\bullet } + i_{R_{E}Y}L_{X}\omega ^{\bullet } + i_{[X , Y]}\omega ^{\bullet \bullet } = \nonumber \\i_{Y}L_{X}\omega ^{\bullet \bullet } - L_{X}i_{Y}\omega ^{\bullet \bullet } + i_{[X , Y]}\omega ^{\bullet \bullet } = 0 
\end{eqnarray}
where we used (\ref{eq:e68}) (\ref{eq:e69}), the property of the Lie derivative 
\begin{eqnarray}
L_{X}i_{Y}\omega = i_{Y}L_{X}\omega + i_{[X , Y]}\omega 
\end{eqnarray}
and the relations of the following type 
\begin{eqnarray}
L_{R_{E}X}\omega = di_{R_{E}X}\omega + i_{R_{E}X}d\omega = di_{X}\omega ^{\bullet } = L_{X}\omega ^{\bullet } - i_{X}d\omega ^{\bullet } = L_{X}\omega ^{\bullet } 
\end{eqnarray}
So we proved that for arbitrary vector fields $X, Y$ the contraction of $T(R_{E})(X , Y)$ and $\omega $ vanishes. But since $W$ bivector is non-degenerate ($W^{n} \neq 0$), its counter image 
\begin{eqnarray}
\omega = \Phi _{\omega }(W)
\end{eqnarray}
is also non-degenerate and vanishing of the contraction (\ref{eq:e70}) implies that the torsion $T(R_{E})$ itself is zero. So we get 
\begin{eqnarray}
T(R_{E})(X , Y) = [R_{E}(X) , R_{E}(Y)] - R_{E}([R_{E}(X) , Y] + [X , R_{E}(Y)] - \nonumber \\
R_{E}([X , Y])) = 0 
\end{eqnarray}
\\
{\bf Sample.} Note that operator $R_{E}$ associated with non-Noether symmetry (\ref{eq:e27}) reproduces well known Fr\"{o}licher-Nijenhuis operator 
\begin{eqnarray}
R_{E} = z_{1}dz_{1} \otimes \frac{ \partial }{\partial z_{1} } - dz_{1} \otimes \frac{ \partial }{\partial z_{4} } + \nonumber \\z_{2}dz_{2} \otimes \frac{ \partial }{\partial z_{2} } + dz_{2} \otimes \frac{ \partial }{\partial z_{3} } + \nonumber \\z_{1}dz_{3} \otimes \frac{ \partial }{\partial z_{3} } + e^{z_{3} - z_{4}}dz_{3} \otimes \frac{ \partial }{\partial z_{2} } + \nonumber \\z_{2}dz_{4} \otimes \frac{ \partial }{\partial z_{4} } - e^{z_{3} - z_{4}}dz_{4} \otimes \frac{ \partial }{\partial z_{1} } 
\end{eqnarray}
(compare with \cite{r30}). Note that operator $\overline{R}_{E}$ plays the role of recursion operator for conservation laws 
\begin{eqnarray}
I^{(1)} = z_{1} + z_{2}\nonumber \\I^{(2)} = z_{1}^{2} + z_{2}^{2} + 2e^{z_{3} - z_{4}} 
\end{eqnarray}
Indeed one can check that 
\begin{eqnarray}
2\overline{R}_{E}(dI^{(1)}) = dI^{(2)}
\end{eqnarray}
Similarly using non-Noether symmetry (\ref{eq:e32}) one can construct recursion operator of three particle Toda chain 
\begin{eqnarray}
R_{E} = z_{1}dz_{1} \otimes \frac{ \partial }{\partial z_{1} } - e^{z_{4} - z_{5}}dz_{5} \otimes \frac{ \partial }{\partial z_{1} }\nonumber \\+ z_{2}dz_{2} \otimes \frac{ \partial }{\partial z_{2} } + e^{z_{4} - z_{5}}dz_{4} \otimes \frac{ \partial }{\partial z_{2} } - e^{z_{5} - z_{6}}dz_{6} \otimes \frac{ \partial }{\partial z_{2} }\nonumber \\+ z_{3}dz_{3} \otimes \frac{ \partial }{\partial z_{3} } + e^{z_{5} - z_{6}}dz_{5} \otimes \frac{ \partial }{\partial z_{3} }\nonumber \\+ z_{1}dz_{4} \otimes \frac{ \partial }{\partial z_{4} } - dz_{2} \otimes \frac{ \partial }{\partial z_{4} } - dz_{3} \otimes \frac{ \partial }{\partial z_{4} }\nonumber \\+ z_{2}dz_{5} \otimes \frac{ \partial }{\partial z_{5} } + dz_{1} \otimes \frac{ \partial }{\partial z_{5} } - dz_{3} \otimes \frac{ \partial }{\partial z_{5} } \nonumber \\+ z_{3}dz_{6} \otimes \frac{ \partial }{\partial z_{6} } + dz_{1} \otimes \frac{ \partial }{\partial z_{6} } + dz_{2} \otimes \frac{ \partial }{\partial z_{6} } 
\end{eqnarray}
and as in case of two particle Toda chain, operator $\overline{R}_{E}$ appears to be recursion operator for conservation laws 
\begin{eqnarray}
I^{(1)} = z_{1} + z_{2}\nonumber \\I^{(2)} = z_{1}^{2} + z_{2}^{2} + z_{3}^{2} + 2e^{z_{4} - z_{5}} + 2e^{z_{5} - z_{6}}\nonumber \\I^{(3)} = z_{1}^{3} + z_{2}^{3} + z_{3}^{3} + 3(z_{1} + z_{2})e^{z_{4} - z_{5}} + 3(z_{2} + z_{3})e^{z_{5} - z_{6}} 
\end{eqnarray}
and fulfills the following recursion condition 
\begin{eqnarray}
dI^{(3)} = 3\overline{R}_{E}(dI^{(2)}) = 6(\overline{R}_{E})^{2}(dI^{(1)})
\end{eqnarray}
\\
\section{One-parameter families of conservation laws}
One-parameter group of transformations $g_{a}$ defined by (\ref{eq:e12}) naturally acts on algebra of integrals of motion. Namely for each conservation law 
\begin{eqnarray}
\frac{ d }{dt }J = 0 
\end{eqnarray}
one can define one-parameter family of conserved quantities $J(a)$ by applying group of transformations $g_{a}$ to $J$ 
\begin{eqnarray}
J(a) = g_{a}(J) = e^{aL_{E}}J = J + aL_{E}J + \frac{ 1 }{2 } (aL_{E})^{2}J + ... 
\end{eqnarray}
Property (\ref{eq:e13}) ensures that $J(a)$ is conserved for arbitrary values of parameter $a$ 
\begin{eqnarray}
\frac{ d }{dt }J(a) = \frac{ d }{dt }g_{a}(J) = g_{a} (\frac{ d }{dt }J) = 0 
\end{eqnarray}
and thus each conservation law gives rise to whole family of conserved quantities that form orbit of group of transformations $g_{a}$. \\
Such an orbit $J(a)$ is called involutive if conservation laws that form it are in involution 
\begin{eqnarray}
\{J(a) , J(b)\} = 0 
\end{eqnarray}
(for arbitrary values of parameters $a, b$). On $2n$ dimensional symplectic manifold each involutive family that contains $n$ functionally independent integrals of motion naturally gives rise to integrable system (due to Liouville-Arnold theorem). So in order to identify those orbits that may be related to integrable models it is important to know how involutivity of family of conserved quantities $J(a)$ is related to properties of initial conserved quantity $J(0) = J$ and nature of generator $E$ of group $g_{a} = e^{aL_{E}}$. In other words we would like to know what condition must be satisfied by generator of symmetry $E$ and integral of motion $J$ to ensure that $\{J(a) , J(b)\} = 0$. To address this issue and to describe class of vector fields that possess nontrivial involutive orbits we would like to propose the following theorem \\
{\bf Theorem 8.} Let $M$ be Poisson manifold endowed with 1-form $s$ such that 

\begin{eqnarray}
\label{eq:e71}
[W[W(s),W](s)] = c_{0}[W(s)[W(s) ,W]] ~~~~~ (c_{0} \neq - 1) 
\end{eqnarray}
Then each function $J$ satisfying property 

\begin{eqnarray}
\label{eq:e72}
W(L_{W(s)}dJ) = c_{1}[W(s),W](dJ)~~~~~ (c_{1} \neq 0) 
\end{eqnarray}
($c_{0,1}$ are some constants) gives rise to involutive set of functions 

\begin{eqnarray}
\label{eq:e73}
J^{(m)} = (L_{W(s)})^{m}J ~~~~~ \{J^{(m)}, J^{(k)}\} = 0 
\end{eqnarray}
\\
{\bf Proof.} First let us inroduce linear operator $R$ on bundle of multivector fields and define it for arbitrary multivector field $V$ by condition 

\begin{eqnarray}
\label{eq:e74}
R(V) = \frac{ 1 }{2 } ([W(s),V] - \Phi _{W}(L_{W(s)}\Phi _{\omega }(V))) 
\end{eqnarray}
Proof of linearity of this operator is identical to proof given for (\ref{eq:e40}) so we will skip it. Further it is clear that 

\begin{eqnarray}
\label{eq:e75}
R(W) = [W(s),W] 
\end{eqnarray}
and 

\begin{eqnarray}
\label{eq:e76}
R^{2}(W) = R([W(s),W]) = \frac{ 1 }{2 }([W(s)[W(s),W]] - \Phi _{W}((L_{W(s)})^{2}\omega )) = \nonumber \\\frac{ 1 + c_{0} }{2 } [W(s)[W(s),W]] 
\end{eqnarray}
where we used property 
\begin{eqnarray}
\Phi _{W}((L_{W(s)})^{2}\omega ) = 
\Phi _{W}(L_{W(s)}L_{W(s)}\omega ) = \nonumber \\
\Phi _{W}(i_{W(s)}dL_{W(s)}\omega ) + 
\Phi _{W}(di_{W(s)}L_{W(s)}\omega ) = 
[W,\Phi _{W}(i_{W(s)}L_{W(s)}\omega )] \nonumber \\
= [W[W(s),W](s)] 
= c_{0}[W(s)[W(s),W]] 
\end{eqnarray}
In the same time by taking Lie derivative of (\ref{eq:e75}) along the vector field $W(s)$ one gets 

\begin{eqnarray}
\label{eq:e77}
[W[W(s),W](s)] = (L_{W(s)}R + R^{2})(W) 
\end{eqnarray}
comparing (\ref{eq:e76}) and (\ref{eq:e77}) yields 
\begin{eqnarray}
(1 + c_{0})(L_{W(s)}R + R^{2}) = 2R^{2}
\end{eqnarray}
and thus 

\begin{eqnarray}
\label{eq:e78}
L_{W(s)}R = \frac{ 1 - c_{0} }{1 + c_{0} } R^{2} 
\end{eqnarray}
Further let us rewrite condition (\ref{eq:e72}) as follows 

\begin{eqnarray}
\label{eq:e79}
W(L_{W(s)}dJ) = c_{1}R(W)(dJ) 
\end{eqnarray}
due to linearity of operator $R$ this condition can be extended to 

\begin{eqnarray}
\label{eq:e80}
R^{m}(W)(L_{W(s)}dJ) = c_{1}R^{m + 1}(W)(dJ) 
\end{eqnarray}
Now assuming that the following condition is true 

\begin{eqnarray}
\label{eq:e81}
W((L_{W(s)})^{m}dJ) = c_{m}R^{m}(W)(dJ) 
\end{eqnarray}
let us take its Lie derivative along vector field $W(s)$. We get 

\begin{eqnarray}
\label{eq:e82}
R(W)((L_{W(s)})^{m}dJ) + W((L_{W(s)})^{m + 1}dJ) = \nonumber \\mc_{m}\frac{ 1 - c_{0} }{1 + c_{0} }R^{m + 1}(W)(dJ) + c_{m}R^{m + 1}(W)(dJ) 
\end{eqnarray}
where we used properties (\ref{eq:e75}) and (\ref{eq:e78}). Note also that (\ref{eq:e81}) together with linearity of operator $R$ imply that 

\begin{eqnarray}
\label{eq:e83}
R^{k}W((L_{W(s)})^{m}dJ) = c_{m}R^{k + m}(W)(dJ) 
\end{eqnarray}
and thus (\ref{eq:e82}) reduces to 

\begin{eqnarray}
\label{eq:e84}
W((L_{W(s)})^{m + 1}dJ) = c_{m + 1}R^{m + 1}(W)(dJ) 
\end{eqnarray}
where $c_{m + 1}$ is defined by 
\begin{eqnarray}
c_{m + 1} = mc_{n}\frac{ 1 - c_{0} }{1 + c_{0} }
\end{eqnarray}
So we proved that if assumtion (\ref{eq:e81}) is valid for $m$ then it is also valid for $m + 1$, we also know that for $m = 1$ it matches (\ref{eq:e79}) and thus by induction we proved that condition (\ref{eq:e81}) is valid for arbitrary $m$ while $c_{n}$ can be determined by 
\begin{eqnarray}
c_{m} = c_{0}(m - 1)! \left [\frac{ 1 - c_{0} }{1 + c_{0} }\right ]^{m - 1} 
\end{eqnarray}
Now using (\ref{eq:e81}) and (\ref{eq:e83}) it is easy to show that functions $(L_{W(s)})^{m}J$ are in involution. Indeed 
\begin{eqnarray}
\{(L_{W(s)})^{m}J, (L_{W(s)})^{k}J\} = W(d(L_{W(s)})^{m}J \wedge d(L_{W(s)})^{k}J) = \nonumber \\W((L_{W(s)})^{m}dJ \wedge (L_{W(s)})^{k}dJ) = c_{m}c_{k}W(dJ \wedge dJ) = 0 
\end{eqnarray}
So we proved functions (\ref{eq:e73}) are in involution. \\
Further we will focus on concrete integrable models such as Toda chain, Broer-Kaup system and Benney system and we will use this theorem to prove involutivity of families of conservation laws constructed using non-Noether symmetries. \\
\section{Toda Model}
To illustrate features of non-Noether symmetries we often refer to two and three particle non-periodic Toda systems. However it turns out that non-Noether symmetries are present in generic n-particle non-periodic Toda chains too, moreover they preserve basic features of symmetries (\ref{eq:e27}), (\ref{eq:e32}). In case of n-particle Toda model symmetry yields $n$ functionally independent conservation laws in involution, gives rise to bi-Hamiltonian structure of Toda hierarchy, reproduces Lax pair of Toda system, endows phase space with Fr\"{o}licher-Nijenhuis operator and leads to invariant bidifferential calculus on algebra of differential forms over phase space of Toda system.\\
First of all let us remind that Toda model is $2n$ dimensional Hamiltonian system that describes the motion of $n$ particles on the line governed by the exponential interaction. Equations of motion of the non periodic n-particle Toda model are 

\begin{eqnarray}
\label{eq:e85}
\frac{ d }{dt }q_{i} = p_{i}\nonumber \\\frac{ d }{dt }p_{i} = \epsilon (i - 1)e^{q_{i - 1} - q_{i}} - \epsilon (n - i)e^{q_{i} - q_{i + 1}} 
\end{eqnarray}
($\epsilon (k) = - \epsilon (- k) = 1$ for any natural $k$ and $\epsilon (0) = 0$) and can be rewritten in Hamiltonian form (\ref{eq:e11}) with canonical Poisson bracket defined by 
\begin{eqnarray}
W = \sum ^{ n }_{i = 1 } \frac{ \partial }{\partial p_{i} } \wedge \frac{ \partial }{\partial q_{i} } 
\end{eqnarray}
corresponding symplectic form 
\begin{eqnarray}
\omega = \sum ^{ n }_{i = 1 }dp_{i} \wedge dq_{i} 
\end{eqnarray}
and Hamiltonian equal to 
\begin{eqnarray}
h = \frac{ 1 }{2 }\sum ^{ n }_{i = 1 }p_{i}^{2} + \sum ^{ n - 1 }_{i = 1 }e^{q_{i} - q_{i + 1}} 
\end{eqnarray}
Note that in two and three particle case we used slightly different notations 
\begin{eqnarray}
z_{i} = p_{i} \nonumber \\z_{n + i} = q_{i} ~~~~~ i = 1, 2, (3); n = 2(3) 
\end{eqnarray}
for local coordinates. The group of transformations $g_{a}$ generated by the vector field $E$ will be symmetry of Toda chain if for each $p_{i}, q_{i}$ satisfying Toda equations (\ref{eq:e85}) $g_{a}(p_{i}), g_{a}(q_{i})$ also satisfy it. Substituting infinitesimal transformations 
\begin{eqnarray}
g_{a}(p_{i}) = p_{i} + aE(p_{i}) + O(a^{2})\nonumber \\g_{a}(p_{i}) = q_{i} + aE(q_{i}) + O(a^{2}) 
\end{eqnarray}
into (\ref{eq:e85}) and grouping first order terms gives rise to the conditions 

\begin{eqnarray}
\label{eq:e86}
\frac{ d }{dt }E(q_{i}) = E(p_{i})\nonumber \\\frac{ d }{dt }E(p_{i}) = \epsilon (i - 1)e^{q_{i - 1} - q_{i}} (E(q_{i - 1}) - E(q_{i})) - \nonumber \\\epsilon (n - i)e^{q_{i} - q_{i + 1}} (E(q_{i}) - E(q_{i + 1})) 
\end{eqnarray}
One can verify that the vector field defined by 

\begin{eqnarray}
\label{eq:e87}
E(p_{i}) = \frac{ 1 }{2 }p_{i}^{2} + \epsilon (i - 1)(n - i + 2) e^{q_{i - 1} - q_{i}} - \epsilon (n - i)(n - i) e^{q_{i} - q_{i + 1}} +\nonumber \\\frac{ t }{2 }(\epsilon (i - 1)(p_{i - 1} + p_{i}) e^{q_{i - 1} - q_{i}} - \epsilon (n - i)(p_{i} + p_{i + 1}) e^{q_{i} - q_{i + 1}} \nonumber \\E(q_{i}) = (n - i + 1)p_{i} - \frac{ 1 }{2 }\sum ^{ i - 1 }_{k = 1 }p_{k} + \frac{ 1 }{2 }\sum ^{ n }_{k = i + 1 }p_{k} + \nonumber \\\frac{ t }{2 }(p_{i}^{2} + \epsilon (i - 1)e^{q_{i - 1} - q_{i}} + \epsilon (n - i)e^{q_{i} - q_{i + 1}}) 
\end{eqnarray}
satisfies (\ref{eq:e15}) and generates symmetry of Toda chain. It appears that this symmetry is non-Noether since it does not preserve Poisson bracket structure $[E , W] \neq 0$ and additionally one can check that Yang-Baxter equation $[[E[E , W]]W] = 0$ is satisfied. This symmetry may play important role in analysis of Toda model. First let us note that calculating $L_{E}\omega $ leads to the following 2-form 

\begin{eqnarray}
\label{eq:e88}
L_{E}\omega = \sum ^{ n }_{i = 1 }p_{i}dp_{i} \wedge dq_{i} + \sum ^{ n - 1 }_{i = 1 }e^{q_{i} - q_{i + 1}} dq_{i} \wedge dq_{i + 1} + \sum ^{ ~ }_{i < j }dp_{i} \wedge dp_{j} 
\end{eqnarray}
and together $\omega $ and $L_{E}\omega $ give rise to bi-Hamiltonian structure of Toda model (compare with \cite{r30}). Thus bi-Hamiltonian realization of Toda chain can be considered as manifestation of hidden symmetry. In fact non-Noether symmetry carries even more information about the bi-Hamiltonian structure and give rise to the invariant symplectic potential for the differential form $L_{E}\omega $, i.e. invariant 1-form $\theta ^{\bullet }$ such that 
\begin{eqnarray}
d\theta ^{\bullet } = \omega 
\end{eqnarray}
This 1-form can be constructed by taking contraction of generator $E$ of non-Noether symmetry and symplectic form $\omega $ 

\begin{eqnarray}
\label{eq:e89}
\theta ^{\bullet } = i_{E}\omega = \sum ^{ n }_{i = 1 } \left [\frac{1}{2}p_{i}^{2}dq_{i} + (n - i + 1)p_{i}dp_{i}\right ] +\nonumber \\\sum ^{ n - 1 }_{i = 1 }e^{q_{i} - q_{i + 1}} \left [(n - i + 1)dq_{i + 1} - (n - i)dq_{i}\right ] + \frac{ 1 }{2 } \sum ^{ ~ }_{i < j } (p_{i}dp_{j} - p_{j}dp_{i}) 
\end{eqnarray}
In terms of bivector fields these bi-Hamiltonian system is formed by 
\begin{eqnarray}
W = \sum ^{ n }_{i = 1 } \frac{ \partial }{\partial p_{i} } \wedge \frac{ \partial }{\partial q_{i} } 
\end{eqnarray}
and 
\begin{eqnarray}
\hat{W} = [E , W] = \sum ^{ n }_{i = 1 }p_{i} \frac{ \partial }{\partial p_{i} } \wedge \frac{ \partial }{\partial q_{i} } + \sum ^{ n - 1 }_{i = 1 }e^{q_{i} - q_{i + 1}} \frac{ \partial }{\partial p_{i} } \wedge \frac{ \partial ~~ }{\partial p_{i + 1} } + \sum ^{ ~ }_{i < j } \frac{ \partial }{\partial q_{i} } \wedge \frac{ \partial }{\partial q_{j} } 
\end{eqnarray}
The conservation laws (\ref{eq:e22}) associated with the symmetry reproduce well known set of conservation laws of Toda chain. 

\begin{eqnarray}
\label{eq:e90}
I^{(1)} = C^{(1)} = \sum ^{ n }_{i = 1 }p_{i}\nonumber \\I^{(2)} = (C^{(1)})^{2} - 2C^{(2)} = \sum ^{ n }_{i = 1 }p_{i}^{2} + 2\sum ^{ n - 1 }_{i = 1 }e^{q_{i} - q_{i + 1}}\nonumber \\I^{(3)} = C^{(1)})^{3} - 3C^{(1)}C^{(2)} + 3C^{(3)} = \sum ^{ n }_{i = 1 }p_{i}^{3} + 3\sum ^{ n - 1 }_{i = 1 }(p_{i} + p_{i + 1}) e^{q_{i} - q_{i + 1}}\nonumber \\I^{(4)} = C^{(1)})^{4} - 4(C^{(1)})^{2}C^{(2)} + 2(C^{(2)})^{2} + 4C^{(1)}C^{(3)} - 4C^{(4)} = \nonumber \\\sum ^{ n }_{i = 1 }p_{i}^{4} + 4\sum ^{ n - 1 }_{i = 1 }(p_{i}^{2} + 2p_{i}p_{i + 1} + p_{i + 1}^{2}) e^{q_{i} - q_{i + 1}} +\nonumber \\2\sum ^{ n - 1 }_{i = 1 }e^{2(q_{i} - q_{i + 1})} + 4\sum ^{ n - 2 }_{i = 1 }e^{q_{i} - q_{i + 2}} \nonumber \\I^{(m)} = (- 1)^{m + 1}mC^{(m)} + \sum ^{ m - 1 }_{k = 1 }(- 1)^{k + 1}I^{(m - k)}C^{(k)} 
\end{eqnarray}
The condition $[[E[E , W]]W] = 0$ satisfied by generator of the symmetry $E$ ensures that the conservation laws are in involution i. e. $\{C^{(k)} , C^{(m)}\} = 0$. Thus the conservation laws as well as the bi-Hamiltonian structure of the non periodic Toda chain appear to be associated with non-Noether symmetry.\\
Using formula (\ref{eq:e39}) one can calculate Lax pair associated with symmetry (\ref{eq:e87}). Lax matrix calculated in this way has the following non-zero entries (note that in case of $n = 2$ and $n = 3$ this formula yields matrices (\ref{eq:e44})-(\ref{eq:e47})) 

\begin{eqnarray}
\label{eq:e91}
L_{k, k} = L_{n + k, n + k} = p_{k}\nonumber \\L_{n + k, k + 1} = - L_{n + k + 1, k} = \epsilon (n - k)e^{q_{k} - q_{k + 1}}\nonumber \\L_{k, n + m} = \epsilon (m - k)\nonumber \\m, k = 1, 2, ... , n 
\end{eqnarray}
while non-zero entries of $P$ matrix involved in Lax pair are 

\begin{eqnarray}
\label{eq:e92}
P_{k, n + k} = 1\nonumber \\P_{n + k, k} = - \epsilon (k - 1)e^{q_{k - 1} - q_{k}} - \epsilon (n - k)e^{q_{k} - q_{k + 1}}\nonumber \\P_{n + k, k + 1} = \epsilon (n - k)e^{q_{k} - q_{k + 1}}\nonumber \\P_{n + k, k - 1} = \epsilon (k - 1)e^{q_{k - 1} - q_{k}}\nonumber \\k = 1, 2, ... , n 
\end{eqnarray}
This Lax pair constructed from generator of non-Noether symmetry exactly reproduces known Lax pair of Toda chain. \\
Like two and three particle Toda chain, n-particle Toda model also admits invariant bidifferential calculus on algebra of differential forms over the phase space. This bidifferential calculus can be constructed using non-Noether symmetry (see (\ref{eq:e63})), it consists out of two differential operators $d, \bar{d}$ where $d$ is ordinary exterior derivative while $\bar{d}$ can be defined by 

\begin{eqnarray}
\label{eq:e93}
\bar{d}q_{i} = p_{i}dq_{i} + \sum ^{ ~ }_{i < j }dp_{j} - \sum ^{ ~ }_{i > j }dp_{j}\nonumber \\\bar{d}p_{i} = p_{i}dp_{i} - e^{q_{i} - q_{i + 1}}dq_{i + 1} + e^{q_{i - 1} - q_{i}}dq_{i} 
\end{eqnarray}
and is extended to whole De Rham complex by linearity, derivation property and compatibility property $d\bar{d} + \bar{d}d = 0$. By direct calculations one can verify that calculus constructed in this way is consistent and satisfies $\bar{d}^{2} = 0$ property. One can also check that conservation laws (\ref{eq:e90}) form Lenard scheme 
\begin{eqnarray}
(k + 1)\bar{d}I^{(k)} = kdI^{(k + 1)}
\end{eqnarray}
\\
Further let us focus on Fr\"{o}licher-Nijenhuis geometry. Using formula (\ref{eq:e66}) one can construct invariant Fr\"{o}licher-Nijenhuis operator, out of generator of non-Noether symmetry of Toda chain. Operator constructed in this way has the form 

\begin{eqnarray}
\label{eq:e94}
\overline{R}_{E} = \sum ^{ n }_{i = 1 }p_{i} \left [ dp_{i} \otimes \frac{ \partial }{\partial q_{i} } + dq_{i} \otimes \frac{ \partial }{\partial p_{i} } \right ] - \sum ^{ n - 1 }_{i = 1 } e^{q_{i} - q_{i + 1}}dq_{i + 1} \otimes \frac{ \partial }{\partial p_{i} }\nonumber \\+ \sum ^{ n - 1 }_{i = 1 } e^{q_{i - 1} - q_{i}}dq_{i} \otimes \frac{ \partial }{\partial p_{i} } - \sum ^{ ~ }_{i < j } \left [ dp_{i} \otimes \frac{ \partial }{\partial q_{j} } - dp_{j} \otimes \frac{ \partial }{\partial q_{i} } \right ] 
\end{eqnarray}
One can check that Fr\"{o}licher-Nijenhuis torsion of this operator vanishes and it plays role of recursion operator for n-particle Toda chain in sense that conservation laws $I^{(k)}$ satisfy recursion relation 

\begin{eqnarray}
\label{eq:e95}
(k + 1)R_{E}(dI^{(k)}) = kdI^{(k + 1)}
\end{eqnarray}
Thus non-Noether symmetry of Toda chain not only leads to n functionally independent conservation laws in involution, but also essentially enriches phase space geometry by endowing it with invariant Fr\"{o}licher-Nijenhuis operator, bi-Hamiltonian system, bicomplex structure and Lax pair. \\
Finally, in order to outline possible applications of Theorem 8 let us study action of non-Noether symmetry (\ref{eq:e87}) on conserved quantities of Toda chain. Vector field $E$ defined by (\ref{eq:e87}) generates one-parameter group of transformations (\ref{eq:e12}) that maps arbitrary conserved quantity $J$ to 
\begin{eqnarray}
J(a) = J + aJ^{(1)} + \frac{ a^{2} }{2! }J^{(2)} + \frac{ a^{3} }{3! }J^{(3)} + ... 
\end{eqnarray}
where 
\begin{eqnarray}
J^{(m)} = (L_{E})^{m}J 
\end{eqnarray}
In particular let us focus on family of conserved quantities obtained by action of $g_{a} = e^{aL_{E}}$ on total momenta of Toda chain 

\begin{eqnarray}
\label{eq:e96}
J = \sum ^{ n }_{i = 1 }p_{i}
\end{eqnarray}
By direct calculations one can check that family $J(a)$, that forms orbit of non-Noether symmetry generated by $E$, reproduces entire involutive family of integrals of motion (\ref{eq:e90}). Namely 

\begin{eqnarray}
\label{eq:e97}
J^{(1)} = L_{E}J = \frac{ 1 }{2 } \sum ^{ n }_{i = 1 }p_{i}^{2} + \sum ^{ n - 1 }_{i = 1 }e^{q_{i} - q_{i + 1}}\nonumber \\J^{(2)} = L_{E}J^{(1)} = (L_{E})^{2}J = \nonumber \\\frac{ 1 }{2 } \sum ^{ n }_{i = 1 }p_{i}^{3} + \frac{ 3 }{2 } \sum ^{ n - 1 }_{i = 1 }(p_{i} + p_{i + 1}) e^{q_{i} - q_{i + 1}}\nonumber \\J^{(3)} = L_{E}J^{(2)} = (L_{E})^{3}J = \nonumber \\\frac{ 3 }{4 } \sum ^{ n }_{i = 1 }p_{i}^{4} + 3\sum ^{ n - 1 }_{i = 1 }(p_{i}^{2} + 2p_{i}p_{i + 1} + p_{i + 1}^{2}) e^{q_{i} - q_{i + 1}} +\nonumber \\\frac{ 3 }{2 } \sum ^{ n - 1 }_{i = 1 }e^{2(q_{i} - q_{i + 1})} + 3\sum ^{ n - 2 }_{i = 1 }e^{q_{i} - q_{i + 2}} \nonumber \\J^{(m)} = L_{E}J^{(m - 1)} = (L_{E})^{m}J 
\end{eqnarray}
\\
Involutivity of this set of conservation laws can be verified using Theorem 8. In particular one can notice that differential 1-form $s$ defined by 
\begin{eqnarray}
E = W(s)
\end{eqnarray}
(where $E$ is generator of non-Noether symmetry (\ref{eq:e87})) satisfies condition 
\begin{eqnarray}
[W[W(s),W](s)] = 3[W(s)[W(s) ,W]] 
\end{eqnarray}
while conservation law $J$ defined by (\ref{eq:e96}) has property 
\begin{eqnarray}
W(L_{W(s)}dJ) = - [W(s),W](dJ) 
\end{eqnarray}
and thus according to Theorem 8 conservation laws (\ref{eq:e97}) are in involution. \\
\section{Nonlinear Schr\"{o}dinger equation}
Toda model provided good sample of finite dimensional integrable Hamiltonian system that possesses non-Noether symmetry. However there are many infinite dimensional integrable Hamiltonian systems and in this case in order to ensure integrability one should construct infinite number of conservation laws. Fortunately in several integrable models this task can be effectively simplified by identifying appropriate non-Noether symmetry. First let us consider well known infinite dimensional integrable Hamiltonian system -- nonlinear Schr\"{o}dinger equation (NSE) 

\begin{eqnarray}
\label{eq:e98}
\psi _{t} = i(\psi _{xx} + 2\psi ^{2}\overline{\psi })
\end{eqnarray}
where $\psi $ is a smooth complex function of $(t, x) \in R^{2}$. On this stage we will not specify any boundary conditions and will just focus on symmetries of NSE. Supposing that the vector field $E$ generates the symmetry of NSE one gets the following restriction 

\begin{eqnarray}
\label{eq:e99}
E(\psi )_{t} = i[E(\psi )_{xx} + 2\psi ^{2}E(\overline{\psi }) + 4\psi \overline{\psi }E(\psi )]
\end{eqnarray}
(obtained by substituting infinitesimal transformation $\psi \rightarrow \psi + aE(\psi ) + O(a^{2})$ generated by $E$ into NSE). It appears that NSE possesses nontrivial symmetry that is generated by the vector field 

\begin{eqnarray}
\label{eq:e100}
E(\psi ) = i(\psi _{x} + \frac{ x }{2 }\psi _{xx} + \psi \phi + x\psi ^{2}\overline{\psi }) - t(\psi _{xxx} + 6\psi \overline{\psi }\psi _{x})
\end{eqnarray}
(here $\phi $ is defined by $\phi _{x} = \psi \overline{\psi }$).\\
In order to construct conservation laws we also need to know Poisson bracket structure and it appears that invariant Poisson bivector field can be defined if $\psi $ is subjected zero $\psi (t, - \infty ) = \psi (t, + \infty ) = 0$ boundary conditions. In terms of variational derivatives the explicit form of the Poisson bivector field is 

\begin{eqnarray}
\label{eq:e101}
W = i \int ^{ + \infty }_{- \infty }dx \frac{ \delta }{\delta \psi } \wedge \frac{ \delta }{\delta \overline{\psi } }
\end{eqnarray}
while corresponding symplectic form obtained by inverting $W$ is 

\begin{eqnarray}
\label{eq:e102}
\omega = i \int ^{ + \infty }_{- \infty }dx \delta \psi \wedge \delta \overline{\psi } 
\end{eqnarray}
Now one can check that NSE can be rewritten in Hamiltonian form 

\begin{eqnarray}
\label{eq:e103}
\psi _{t} = \{h , \psi \}
\end{eqnarray}
with Poisson bracket $\{ , \}$ defined by $W$ and 

\begin{eqnarray}
\label{eq:e104}
h = \int ^{ + \infty }_{- \infty }dx (\psi ^{2}\overline{\psi }^{2} - \psi _{x}\overline{\psi }_{x}) 
\end{eqnarray}
\\
Knowing the symmetry of NSE that appears to be non-Noether ($[E, W] \neq 0$) one can construct bi-Hamiltonian structure and conservation laws. First let us calculate Lie derivative of symplectic form along the symmetry generator 

\begin{eqnarray}
\label{eq:e105}
L_{E}\omega = \int ^{ + \infty }_{- \infty }[\delta \psi _{x} \wedge \delta \overline{\psi } + \psi \delta \phi \wedge \delta \overline{\psi } + \overline{\psi }\delta \phi \wedge \delta \psi ]dx 
\end{eqnarray}
The couple of 2-forms $\omega $ and $L_{E}\omega $ exactly reproduces the bi-Hamiltonian structure of NSE proposed by Magri \cite{r55}. Note also that using non-Noether symmetry one can construct invariant symplectic potential 

\begin{eqnarray}
\label{eq:e106}
\theta ^{\bullet } = i_{E}\omega = \int ^{ + \infty }_{- \infty } [\frac{ 1 }{2 } (\psi \delta \overline{\psi }_{x} + \overline{\psi }\delta \psi _{x}) + \phi \delta (\psi \overline{\psi }) + \nonumber \\\frac{ x }{2 } (\psi ^{2}\overline{\psi }^{2} - \psi _{x}\overline{\psi }_{x})]dx + t\delta [\int ^{ + \infty }_{- \infty }(\overline{\psi }_{x}\psi - \psi _{x}\overline{\psi }) dx] 
\end{eqnarray}
\\
The the conservation laws associated with non-Noether symmetry are well known conservation laws of NSE 

\begin{eqnarray}
\label{eq:e107}
I^{(1)} = C^{(1)} = 2 \int ^{ + \infty }_{- \infty }\psi \overline{\psi } dx\nonumber \\I^{(2)} = (C^{(1)})^{2} - 2C^{(2)} = i \int ^{ + \infty }_{- \infty }(\overline{\psi }_{x}\psi - \psi _{x}\overline{\psi }) dx\nonumber \\I^{(3)} = (C^{(1)})^{3} - 3C^{(1)}C^{(2)} + 3C^{(3)} = 2 \int ^{ + \infty }_{- \infty } (\psi ^{2}\overline{\psi }^{2} - \psi _{x}\overline{\psi }_{x}) dx \nonumber \\I^{(4)} = (C^{(1)})^{4} - 4(C^{(1)})^{2}C^{(2)} + 2(C^{(2)})^{2} + 4C^{(1)}C^{(3)} - 4C^{(4)} = \nonumber \\\int ^{ + \infty }_{- \infty }[i(\overline{\psi }_{x}\psi _{xx} - \psi _{x}\overline{\psi }_{xx}) + 3i(\overline{\psi }\psi ^{2}\overline{\psi }_{x} - \psi \overline{\psi }^{2}\psi _{x})] dx \nonumber \\I^{(m)} = (- 1)^{m + 1}mC^{(m)} + \sum ^{ m - 1 }_{k = 1 }(- 1)^{k + 1}I^{(m - k)}C^{(k)} 
\end{eqnarray}
The involutivity of the conservation laws of NSE $\{C^{(k)}, C^{(m)}\} = 0 $ is related to the fact that $E$ satisfies Yang-Baxter equation $[[E[E , W]]W] = 0$. So non-Noether symmetry of NSE reproduces infinite sequence of functionally independent conservation laws in involution, endows the phase space with invariant bi-Hamiltonian structure and gives rise to the following Fr\"{o}licher-Nijenhuis operator 

\begin{eqnarray}
\label{eq:e108}
R_{E} = \int ^{ + \infty }_{- \infty } i[\psi \delta \phi \otimes \frac{ \delta }{\delta \psi } - \frac{ 1 }{2 } \delta \psi _{x} \otimes \frac{ \delta }{\delta \psi }] dx + h. c. 
\end{eqnarray}
This Fr\"{o}licher-Nijenhuis operator plays the role of recursion operator for the infinite sequence of conservation laws 
\begin{eqnarray}
dI^{(k + 1)} = R_{E}(dI^{(k)})
\end{eqnarray}
\\
\section{Korteweg-de Vries equation}
Now let us consider other important integrable models -- Korteweg-de Vries equation (KdV) and modified Korteweg-de Vries equation (mKdV). Here symmetries are more complicated but generator of the symmetry still can be identified and used in construction of conservation laws. The KdV and mKdV equations have the following form 

\begin{eqnarray}
\label{eq:e109}
u_{t} + u_{xxx} + uu_{x} = 0 [KdV]
\end{eqnarray}
and 

\begin{eqnarray}
\label{eq:e110}
u_{t} + u_{xxx} - 6u^{2}u_{x} = 0 [mKdV]
\end{eqnarray}
(here $u$ is smooth function of $(t, x) \in R^{2}$). The generators of symmetries of KdV and mKdV should satisfy conditions 

\begin{eqnarray}
\label{eq:e111}
E(u)_{t} + E(u)_{xxx} + u_{x}E(u) + uE(u)_{x} = 0 [KdV]
\end{eqnarray}
and 

\begin{eqnarray}
\label{eq:e112}
E(u)_{t} + E(u)_{xxx} - 12uu_{x}E(u) - 6u^{2}E(u)_{x} = 0 [mKdV]
\end{eqnarray}
(again this conditions are obtained by substituting infinitesimal transformation $u \rightarrow u + aE(u) + O(a^{2})$ into KdV and mKdV, respectively). \\
Further we will focus on the symmetries generated by the following vector fields 

\begin{eqnarray}
\label{eq:e113}
E(u) = \frac{ 1 }{2 }u_{xx} + \frac{ 1 }{6 }u^{2} + \frac{ 1 }{24 }u_{x}v + \frac{ x }{8 }(u_{xxx} + uu_{x}) - \nonumber \\\frac{ t }{16 }(6u_{xxxxx} + 20u_{x}u_{xx} + 10 uu_{xxx} + 5u^{2}u_{x}) [KdV] 
\end{eqnarray}
and 

\begin{eqnarray}
\label{eq:e114}
E(u) = - \frac{ 3 }{2 }u_{xx} + 2u^{3} + u_{x}w - \frac{ x }{2 }(u_{xxx} - 6u^{2}u_{x}) -\nonumber \\\frac{ 3t }{2 }(u_{xxxxx} - 10u^{2}u_{xxx} - 40uu_{x}u_{xx} - 10u_{x}^{3} + 30u^{4}u_{x}) [mKdV]
\end{eqnarray}
(here $v$ and $w$ are defined by $v_{x} = u$ and $w_{x} = u^{2}$)\\
To construct conservation laws we need to know Poisson bracket structure and again like in the case of NSE the Poisson bivector field is well defined when $u$ is subjected to zero $u(t, - \infty ) = u(t, + \infty ) = 0$ boundary conditions. For both KdV and mKdV the Poisson bivector field is 

\begin{eqnarray}
\label{eq:e115}
W = \int ^{ + \infty }_{- \infty }dx \frac{ \delta }{\delta u }\wedge \frac{ \delta }{\delta v }
\end{eqnarray}
with corresponding symplectic form 

\begin{eqnarray}
\label{eq:e116}
\omega = \int ^{ + \infty }_{- \infty }dx \delta u \wedge \delta v 
\end{eqnarray}
leading to Hamiltonian realization of KdV and mKdV equations 

\begin{eqnarray}
\label{eq:e117}
u_{t} = \{h , u\}
\end{eqnarray}
with Hamiltonians 

\begin{eqnarray}
\label{eq:e118}
h = \int ^{ + \infty }_{- \infty }(u_{x}^{2} - \frac{ u^{3} }{3 }) dx [KdV]
\end{eqnarray}
and 

\begin{eqnarray}
\label{eq:e119}
h = \int ^{ + \infty }_{- \infty }(u_{x}^{2} + u^{4}) dx [mKdV]
\end{eqnarray}
By taking Lie derivative of the symplectic form along the generators of the symmetries one gets another couple of symplectic forms 

\begin{eqnarray}
\label{eq:e120}
L_{E}\omega = \int ^{ + \infty }_{- \infty }dx (\delta u \wedge \delta u_{x} + \frac{ 2 }{3 }u\delta u \wedge \delta v) [KdV] 
\end{eqnarray}
\begin{eqnarray}
L_{E}\omega = \int ^{ + \infty }_{- \infty }dx (\delta u \wedge \delta u_{x} - 2u\delta u \wedge \delta w) [mKdV] 
\end{eqnarray}
involved in bi-Hamiltonian realization of KdV/mKdV hierarchies and proposed by Magri \cite{r55}. \\
The conservation laws associated with the symmetries reproduce infinite sequence of conservation laws of KdV equation 

\begin{eqnarray}
\label{eq:e121}
I^{(1)} = C^{(1)} = \frac{ 2 }{3 }\int ^{ + \infty }_{- \infty }u dx \nonumber \\I^{(2)} = C^{(1)} - 2C^{(2)} = \frac{ 4 }{9 }\int ^{ + \infty }_{- \infty }u^{2} dx \nonumber \\I^{(3)} = (C^{(1)})^{3} - 3C^{(1)}C^{(2)} + 3C^{(3)} = \frac{ 8 }{9 }\int ^{ + \infty }_{- \infty }(\frac{ u^{3} }{3 } - u_{x}^{2}) dx \nonumber \\I^{(4)} = (C^{(1)})^{4} - 4(C^{(1)})^{2}C^{(2)} + 2(C^{(2)})^{2} + 4C^{(1)}C^{(3)} - 4C^{(4)} = \nonumber \\\frac{ 64 }{45 }\int ^{ + \infty }_{- \infty }(\frac{ 5 }{36 }u^{4} - \frac{ 5 }{3 }uu_{x}^{2} + u_{xx}^{2}) dx \nonumber \\I^{(m)} = (- 1)^{m + 1}mC^{(m)} + \sum ^{ m - 1 }_{k = 1 }(- 1)^{k + 1}I^{(m - k)}C^{(k)} 
\end{eqnarray}
and mKdV equation 

\begin{eqnarray}
\label{eq:e122}
I^{(1)} = C^{(1)} = - 4 \int ^{ + \infty }_{- \infty }u^{2} dx \nonumber \\I^{(2)} = C^{(1)} - 2C^{(2)} = 16 \int ^{ + \infty }_{- \infty }(u^{4} + u_{x}^{2}) dx \nonumber \\I^{(3)} = (C^{(1)})^{3} - 3C^{(1)}C^{(2)} + 3C^{(3)} = - 32 \int ^{ + \infty }_{- \infty }(2u^{6} + 10 u^{2}u_{x}^{2} + u_{xx}^{2}) dx \nonumber \\I^{(4)} = (C^{(1)})^{4} - 4(C^{(1)})^{2}C^{(2)} + 2(C^{(2)})^{2} + 4C^{(1)}C^{(3)} - 4C^{(4)} = \nonumber \\\frac{ 256 }{5 }\int ^{ + \infty }_{- \infty }(5 u^{8} + 70u^{4}u_{x}^{2} - 7u_{x}^{4} + 14u^{2}u_{xx}^{2} + u_{xxx}^{2}) dx \nonumber \\I^{(m)} = (- 1)^{m + 1}mC^{(m)} + \sum ^{ m - 1 }_{k = 1 }(- 1)^{k + 1}I^{(m - k)}C^{(k)} 
\end{eqnarray}
The involutivity of these conservation laws is well known and in terms of the symmetry generators it is ensured by conditions $[[E[E , W]]W] = 0$. Thus the conservation laws and bi-Hamiltonian structures of KdV and mKdV hierarchies are related to the non-Noether symmetries of KdV and mKdV equations. Moreover these symmetries seem to be responsible for the existence of well known bi-Hamiltonian structures of KdV and mKdV equations. \\
\section{Nonlinear water wave equations}
Among nonlinear partial differential equations that describe propagation of waves in shallow water there are many remarkable integrable systems. We already discussed case of KdV and mKdV equations, that possess non-Noether symmetries leading to the infinite sequence of conservation laws and bi-Hamiltonian realization of these equations, now let us consider other important water wave systems. It is reasonable to start with dispersive water wave system, since many other models can be obtained from it by reduction. Evolution of dispersive water wave system is governed by the following set of equations 

\begin{eqnarray}
\label{eq:e123}
u_{t} = u_{x}w + uw_{x}\nonumber \\v_{t} = uu_{x} - v_{xx} + 2v_{x}w + 2vw_{x}\nonumber \\w_{t} = w_{xx} - 2v_{x} + 2ww_{x} 
\end{eqnarray}
Each symmetry of this system must satisfy linear equation 
\begin{eqnarray}
E(u)_{t} = (wE(u))_{x} + (uE(w))_{x}\nonumber \\E(v)_{t} = (uE(u))_{x} - E(v)_{xx} + 2(wE(v))_{x} + 2(vE(w))_{x}\nonumber \\E(w)_{t} = E(w)_{xx} - 2E(v)_{x} + 2(wE(w))_{x} 
\end{eqnarray}
obtained by substituting infinitesimal transformations 
\begin{eqnarray}
u \rightarrow u + aE(u) + O(a^{2})\nonumber \\v \rightarrow v + aE(v) + O(a^{2})\nonumber \\w \rightarrow w + aE(w) + O(a^{2}) 
\end{eqnarray}
into equations of motion (\ref{eq:e123}) and grouping first order (in $a$) terms. One of the solutions of this equation yields the following symmetry of dispersive water wave system 

\begin{eqnarray}
\label{eq:e124}
E(u) = uw + x(uw)_{x} + 2t(uw^{2} - 2uv + uw_{x})_{x}\nonumber \\E(v) = \frac{ 3 }{2 }u^{2} + 4vw - 3v_{x} + x(uu_{x} + 2(vw)_{x} - v_{xx})\nonumber \\+ 2t(u^{2}w - uu_{x} - 3v^{2} + 3vw^{2} - 3v_{x}w + v_{xx})_{x}\nonumber \\E(w) = w^{2} + 2w_{x} - 4v + x(2ww_{x} + w_{xx} - 2v_{x})\nonumber \\- 2t(u^{2} + 6vw - w^{3} - 3ww_{x} - w_{xx})_{x} 
\end{eqnarray}
and it is remarkable that this symmetry is local in sense that $E(u)$ in point $x$ depends only on $u$ and its derivatives evaluated in the same point, (this is not the case in KdV, mKdV and NLS equations where symmetries are non local due to presence of non local fields like $v$ defined by $v_{x} = u$ in KdV equation, $w$ defined by $w_{x} = u^{2}$ in mKdV and $\phi $ defined by $\phi _{x} = \psi \overline{\psi }$ in case of nonlinear Scr\"{o}dinger equation). \\
Before we proceed let us note that dispersive water wave system is actually infinite dimensional Hamiltonian dynamical system. Assuming that $u, v$ and $w$ fields are subjected to zero boundary conditions 
\begin{eqnarray}
u(\pm \infty ) = v(\pm \infty ) = w(\pm \infty ) = 0
\end{eqnarray}
it is easy to verify that equations (\ref{eq:e123}) can be represented in Hamiltonian form 
\begin{eqnarray}
u_{t} = \{h , u\}\nonumber \\v_{t} = \{h , v\}\nonumber \\w_{t} = \{h , w\} 
\end{eqnarray}
with Hamiltonian equal to 

\begin{eqnarray}
\label{eq:e125}
h = - \frac{ 1 }{4 } \int ^{ + \infty }_{- \infty } (u^{2}w + 2vw^{2} - 2v_{x}w - 2v^{2})dx 
\end{eqnarray}
and Poisson bracket defined by the following Poisson bivector field 

\begin{eqnarray}
\label{eq:e126}
W = \int ^{ + \infty }_{- \infty } \left [ \frac{ 1 }{2 } \frac{ \delta }{\delta u } \wedge \left [ \frac{ \delta }{\delta u } \right ]_{x} + \frac{ \delta }{\delta v } \wedge \left [ \frac{ \delta }{\delta w } \right ]_{x} \right ] dx 
\end{eqnarray}
Now using our symmetry that appears to be non-Noether, one can calculate second Poisson bivector field involved in the bi-Hamiltonian realization of dispersive water wave system 

\begin{eqnarray}
\label{eq:e127}
\hat{W} = [E , W] = \nonumber \\
- 2 \int ^{ + \infty }_{- \infty } \left [ u \frac{ \delta }{\delta v } \wedge \left [ \frac{ \delta }{\delta u } \right ]_{x} + v \frac{ \delta }{\delta v } \wedge \left [ \frac{ \delta }{\delta v } \right ]_{x} + \left [ \frac{ \delta }{\delta v } \right ]_{x} \wedge \left [ \frac{ \delta }{\delta w } \right ]_{x} + w \frac{ \delta }{\delta v } \wedge \left [ \frac{ \delta }{\delta w } \right ]_{x} + \left [ \frac{ \delta }{\delta w } \right ]_{x} \wedge \frac{ \delta }{\delta w } \right ] dx 
\end{eqnarray}
Note that $\hat{W}$ give rise to the second Hamiltonian realization of the model 
\begin{eqnarray}
u_{t} = \{h^{\bullet } , u\}_{\bullet }\nonumber \\v_{t} = \{h^{\bullet } , v\}_{\bullet }\nonumber \\w_{t} = \{h^{\bullet } , w\}_{\bullet }\nonumber \\
\end{eqnarray}
where 
\begin{eqnarray}
h^{\bullet } = - \frac{ 1 }{4 } \int ^{ + \infty }_{- \infty } (u^{2} + 2vw)dx 
\end{eqnarray}
and $\{ , \}_{\bullet }$ is Poisson bracket defined by bivector field $\hat{W}$. \\
Now let us pay attention to conservation laws. By integrating third equation of dispersive water wave system (\ref{eq:e123}) it is easy to show that 
\begin{eqnarray}
J^{(0)} = \int ^{ + \infty }_{- \infty } wdx 
\end{eqnarray}
is conservation law. Using non-Noether symmetry one can construct other conservation laws by taking Lie derivative of $J^{(0)}$ along the generator of symmetry and in this way entire infinite sequence of conservation laws of dispersive water wave system can be reproduced 

\begin{eqnarray}
\label{eq:e128}
J^{(0)} = \int ^{ + \infty }_{- \infty } wdx \nonumber \\J^{(1)} = L_{E}J^{(0)} = - 2 \int ^{ + \infty }_{- \infty } vdx \nonumber \\J^{(2)} = L_{E}J^{(1)} = (L_{E})^{2}J^{(0)} = - 2 \int ^{ + \infty }_{- \infty } (u^{2} + 2vw)dx \nonumber \\J^{(3)} = L_{E}J^{(2)} = (L_{E})^{3}J^{(0)} = - 6 \int ^{ + \infty }_{- \infty } (u^{2}w + 2vw^{2} - 2v_{x}w - 2v^{2})dx \nonumber \\J^{(4)} = L_{E}J^{(3)} = (L_{E})^{4}J^{(0)} = \nonumber \\- 24 \int ^{ + \infty }_{- \infty } (u^{2}w^{2} + u^{2}w_{x} - 2u^{2}v - 6v^{2}w + 2vw^{3} - 3v_{x}w^{2} - 2v_{x}w_{x})dx \nonumber \\J^{(n)} = L_{E}J^{(n - 1)} = (L_{E})^{n}J^{(0)} 
\end{eqnarray}
Thus conservation laws and bi-Hamiltonian structure of dispersive water wave system can be constructed by means of non-Noether symmetry. \\
To prove involutivity of infinite sequence of conserved quantities (\ref{eq:e128}) one can use Theorem 8. In particular one can check that 1-form $s$ defined via 
\begin{eqnarray}
E = W(s)
\end{eqnarray}
($E$ is generator of non-Noether symmetry (\ref{eq:e124})) satisfies condition 
\begin{eqnarray}
[W[W(s),W](s)] = 3[W(s)[W(s) ,W]] 
\end{eqnarray}
while 
\begin{eqnarray}
J = \int ^{ + \infty }_{- \infty } vdx 
\end{eqnarray}
has property 
\begin{eqnarray}
W(L_{W(s)}dJ) = - [W(s),W](dJ) 
\end{eqnarray}
and as a consequence of Theorem 8 gives rise to involutive family of conserved quantities. \\
Note that symmetry (\ref{eq:e124}) can be used in many other partial differential equations that can be obtained by reduction from dispersive water wave system. In particular one can use it in dispersiveless water wave system, Broer-Kaup system, dispersiveless long wave system, Burger's equation etc. In case of dispersiveless water waves system 

\begin{eqnarray}
\label{eq:e129}
u_{t} = u_{x}w + uw_{x}\nonumber \\v_{t} = uu_{x} + 2v_{x}w + 2vw_{x}\nonumber \\w_{t} = - 2v_{x} + 2ww_{x} 
\end{eqnarray}
symmetry (\ref{eq:e124}) is reduced to 

\begin{eqnarray}
\label{eq:e130}
E(u) = uw + x(uw)_{x} + 2t(uw^{2} - 2uv)_{x}\nonumber \\E(v) = \frac{ 3 }{2 }u^{2} + 4vw + x(uu_{x} + 2(vw)_{x}) + 2t(u^{2}w - 3v^{2} + 3vw^{2})_{x}\nonumber \\E(w) = w^{2} - 4v + x(2ww_{x} - 2v_{x}) - 2t(u^{2} + 6vw - w^{3})_{x} 
\end{eqnarray}
and corresponding conservation laws (\ref{eq:e128}) reduce to 

\begin{eqnarray}
\label{eq:e131}
J^{(0)} = \int ^{ + \infty }_{- \infty } wdx \nonumber \\J^{(1)} = L_{E}J^{(0)} = - 2 \int ^{ + \infty }_{- \infty } vdx \nonumber \\J^{(2)} = L_{E}J^{(1)} = (L_{E})^{2}J^{(0)} = - 2 \int ^{ + \infty }_{- \infty } (u^{2} + 2vw)dx \nonumber \\J^{(3)} = L_{E}J^{(2)} = (L_{E})^{3}J^{(0)} = - 6 \int ^{ + \infty }_{- \infty } (u^{2}w + 2vw^{2} - 2v^{2})dx \nonumber \\J^{(4)} = L_{E}J^{(3)} = (L_{E})^{4}J^{(0)} = \nonumber \\- 24 \int ^{ + \infty }_{- \infty } (u^{2}w^{2} - 2u^{2}v - 6v^{2}w + 2vw^{3})dx \nonumber \\J^{(n)} = L_{E}J^{(n - 1)} = (L_{E})^{n}J^{(0)} 
\end{eqnarray}
\\
Another important integrable model that can be obtained from dispersive water wave system is Broer-Kaup system 

\begin{eqnarray}
\label{eq:e132}
v_{t} = \frac{ 1 }{2 } v_{xx} + v_{x}w + vw_{x}\nonumber \\w_{t} = - \frac{ 1 }{2 } w_{xx} + v_{x} + ww_{x} 
\end{eqnarray}
One can check that symmetry (\ref{eq:e124}) of dispersive water wave system, after reduction, reproduces non-Noether symmetry of Broer-Kaup model 

\begin{eqnarray}
\label{eq:e133}
E(v) = 4vw + 3v_{x} + x(2(vw)_{x} + v_{xx})\nonumber \\+ t(3v^{2} + 3vw^{2} + 3v_{x}w + v_{xx})_{x}\nonumber \\E(w) = w^{2} - 2w_{x} + 4v + x(2ww_{x} - w_{xx} + 2v_{x})\nonumber \\+ t(6vw + w^{3} - 3ww_{x} + w_{xx})_{x} 
\end{eqnarray}
and gives rise to the infinite sequence of conservation laws of Broer-Kaup hierarchy 

\begin{eqnarray}
\label{eq:e134}
J^{(0)} = \int ^{ + \infty }_{- \infty } wdx \nonumber \\J^{(1)} = L_{E}J^{(0)} = 2 \int ^{ + \infty }_{- \infty } vdx \nonumber \\J^{(2)} = L_{E}J^{(1)} = (L_{E})^{2}J^{(0)} = 4 \int ^{ + \infty }_{- \infty } vwdx \nonumber \\J^{(3)} = L_{E}J^{(2)} = (L_{E})^{3}J^{(0)} = 12 \int ^{ + \infty }_{- \infty } (vw^{2} + v_{x}w + v^{2})dx \nonumber \\J^{(4)} = L_{E}J^{(3)} = (L_{E})^{4}J^{(0)} = \nonumber \\24 \int ^{ + \infty }_{- \infty } (6v^{2}w + 2vw^{3} + 3v_{x}w^{2} - 2v_{x}w_{x})dx \nonumber \\J^{(n)} = L_{E}J^{(n - 1)} = (L_{E})^{n}J^{(0)} 
\end{eqnarray}
\\
And exactly like in the dispersive water wave system one can rewrite equations of motion (\ref{eq:e132}) in Hamiltonian form 
\begin{eqnarray}
v_{t} = \{h , v\}\nonumber \\w_{t} = \{h , w\} 
\end{eqnarray}
where Hamiltonian is 
\begin{eqnarray}
h = \frac{ 1 }{2 } \int ^{ + \infty }_{- \infty } (vw^{2} + v_{x}w + v^{2})dx 
\end{eqnarray}
while Poisson bracket is defined by the Poisson bivector field 

\begin{eqnarray}
\label{eq:e135}
W = \int ^{ + \infty }_{- \infty } \left [ \frac{ \delta }{\delta v } \wedge \left [ \frac{ \delta }{\delta w } \right ]_{x} \right ] dx 
\end{eqnarray}
And again, using symmetry (\ref{eq:e133}) one can recover second Poisson bivector field involved in the bi-Hamiltonian realization of Broer-Kaup system by taking Lie derivative of (\ref{eq:e135}) 

\begin{eqnarray}
\label{eq:e136}
\hat{W} = [E , W] = \nonumber \\
- 2 \int ^{ + \infty }_{- \infty } \left [ v \frac{ \delta }{\delta v } \wedge \left [ \frac{ \delta }{\delta v } \right ]_{x} - \left [ \frac{ \delta }{\delta v } \right ]_{x} \wedge \left [ \frac{ \delta }{\delta w } \right ]_{x} + w \frac{ \delta }{\delta v } \wedge \left [ \frac{ \delta }{\delta w } \right ]_{x} + \frac{ \delta }{\delta w } \wedge \left [ \frac{ \delta }{\delta w } \right ]_{x} \right ] dx 
\end{eqnarray}
This bivector field give rise to the second Hamiltonian realization of the Broer-Kaup system 
\begin{eqnarray}
v_{t} = \{h^{\bullet } , v\}_{\bullet }\nonumber \\w_{t} = \{h^{\bullet } , w\}_{\bullet }\nonumber \\
\end{eqnarray}
with 
\begin{eqnarray}
h^{\bullet } = - \frac{ 1 }{4 } \int ^{ + \infty }_{- \infty } vwdx 
\end{eqnarray}
So the non-Noether symmetry of Broer-Kaup system yields infinite sequence of conservation laws of Broer-Kaup hierarchy and endows it with bi-Hamiltonian structure. \\
By suppressing dispersive terms in Broer-Kaup system one reduces it to more simple integarble model --- dispersiveless long wave system 

\begin{eqnarray}
\label{eq:e137}
v_{t} = v_{x}w + vw_{x}\nonumber \\w_{t} = v_{x} + ww_{x} 
\end{eqnarray}
in this case symmetry (\ref{eq:e124}) reduces to more simple non-Noether symmetry 

\begin{eqnarray}
\label{eq:e138}
E(v) = 4vw + 2x(vw)_{x} + 3t(v^{2} + vw^{2})_{x}\nonumber \\E(w) = w^{2} + 4v + 2x(ww_{x} + v_{x}) + t(6vw + w^{3})_{x} 
\end{eqnarray}
while the conservation laws of Broer-Kaup hierarchy reduce to sequence of conservation laws of dispersiveless long wave system 

\begin{eqnarray}
\label{eq:e139}
J^{(0)} = \int ^{ + \infty }_{- \infty } wdx \nonumber \\J^{(1)} = L_{E}J^{(0)} = 2 \int ^{ + \infty }_{- \infty } vdx \nonumber \\J^{(2)} = L_{E}J^{(1)} = (L_{E})^{2}J^{(0)} = 4 \int ^{ + \infty }_{- \infty } vwdx \nonumber \\J^{(3)} = L_{E}J^{(2)} = (L_{E})^{3}J^{(0)} = 12 \int ^{ + \infty }_{- \infty } (vw^{2} + v^{2})dx \nonumber \\J^{(4)} = L_{E}J^{(3)} = (L_{E})^{4}J^{(0)} = 48 \int ^{ + \infty }_{- \infty } (3v^{2}w + vw^{3})dx \nonumber \\J^{(n)} = L_{E}J^{(n - 1)} = (L_{E})^{n}J^{(0)} 
\end{eqnarray}
\\
In the same time bi-Hamitonian structure of Broer-Kaup hierarchy, after reduction gives rise to bi-Hamiltonian structure of dispersiveless long wave system 

\begin{eqnarray}
\label{eq:e140}
W = \int ^{ + \infty }_{- \infty } \left [ \frac{ \delta }{\delta v } \wedge \left [ \frac{ \delta }{\delta w } \right ]_{x} \right ] dx \nonumber \\
\hat{W} = [E , W] = \nonumber \\
- 2 \int ^{ + \infty }_{- \infty } \left [ v \frac{ \delta }{\delta v } \wedge \left [ \frac{ \delta }{\delta v } \right ]_{x} + w \frac{ \delta }{\delta v } \wedge \left [ \frac{ \delta }{\delta w } \right ]_{x} + \frac{ \delta }{\delta w } \wedge \left [ \frac{ \delta }{\delta w } \right ]_{x} \right ] dx 
\end{eqnarray}
\\
Among other reductions of dispersive water wave system one should probably mention Burger's equation 

\begin{eqnarray}
\label{eq:e141}
w_{t} = w_{xx} + ww_{x} 
\end{eqnarray}
However Hamiltonian realization of this equation is unknown (for instance Poisson bivector field of dispersive water wave system (\ref{eq:e126}) vanishes during reduction). \\
\section{Benney system}
Now let us consider another integrable system of nonlinear partial differential equations --- Benney system. Time evolution of this dynamical system is governed by equations of motion 

\begin{eqnarray}
\label{eq:e142}
u_{t} = vv_{x} + 2(uw)_{x}\nonumber \\v_{t} = 2u_{x} + (vw)_{x}\nonumber \\w_{t} = 2v_{x} + 2ww_{x} 
\end{eqnarray}
To determine symmetries of the system one has to look for solutions of linear equation 

\begin{eqnarray}
\label{eq:e143}
E(u)_{t} = (vE(v))_{x} + 2(uE(w))_{x} + 2(wE(u))_{x}\nonumber \\E(v)_{t} = 2E(u)_{x} + (vE(w))_{x} + (wE(v))_{x}\nonumber \\E(w)_{t} = 2E(v)_{x} + 2(wE(w))_{x} 
\end{eqnarray}
obtained by substituting infinitesimal transformations 
\begin{eqnarray}
u \rightarrow u + aE(u) + O(a^{2})\nonumber \\v \rightarrow v + aE(v) + O(a^{2})\nonumber \\w \rightarrow w + aE(w) + O(a^{2}) 
\end{eqnarray}
into equations (\ref{eq:e142}) and grouping first order terms. In particular one can check that the vector field $E$ defined by 

\begin{eqnarray}
\label{eq:e144}
E(u) = 5uw + 2v^{2} + x(2(uw)_{x} + vv_{x}) + 2t(4uv + v^{2}w + 3uw^{2})_{x}\nonumber \\E(v) = vw + 6u + x((vw)_{x} + 2u_{x}) + 2t(4uw + 3v^{2} + vw^{2})_{x}\nonumber \\E(w) = w^{2} + 4v + 2x(ww_{x} + v_{x}) + 2t(w^{3} + 4vw + 4u)_{x} 
\end{eqnarray}
satisfies equation (\ref{eq:e143}) and therefore generates symmetry of Benney system. The fact that this symmetry is local simplifies further calculations.\\
In the same time, it is known fact, that under zero boundary conditions 
\begin{eqnarray}
u(\pm \infty ) = v(\pm \infty ) = w(\pm \infty ) = 0
\end{eqnarray}
Benney equations can be rewritten in Hamiltonian form 
\begin{eqnarray}
u_{t} = \{h , u\}\nonumber \\v_{t} = \{h , v\}\nonumber \\w_{t} = \{h , w\} 
\end{eqnarray}
with Hamiltonian 

\begin{eqnarray}
\label{eq:e145}
h = - \frac{ 1 }{2 } \int ^{ + \infty }_{- \infty } (2uw^{2} + 4uv + v^{2}w)dx 
\end{eqnarray}
and Poisson bracket defined by the following Poisson bivector field 

\begin{eqnarray}
\label{eq:e146}
W = \int ^{ + \infty }_{- \infty } \left [ \frac{ 1 }{2 } \frac{ \delta }{\delta v } \wedge \left [ \frac{ \delta }{\delta v } \right ]_{x} + \frac{ \delta }{\delta u } \wedge \left [ \frac{ \delta }{\delta w } \right ]_{x} \right ] dx 
\end{eqnarray}
Using symmetry (\ref{eq:e144}) that in fact is non-Noether one, we can reproduce second Poisson bivector field involved in the bi-Hamiltonian structure of Benney hierarchy (by taking Lie derivative of $W$ along $E$) 

\begin{eqnarray}
\label{eq:e147}
\hat{W} = [E , W] = \nonumber \\
- 3 \int ^{ + \infty }_{- \infty } \left [ u \frac{ \delta }{\delta u } \wedge \left [ \frac{ \delta }{\delta u } \right ]_{x} + v \frac{ \delta }{\delta u } \wedge \left [ \frac{ \delta }{\delta v } \right ]_{x} + w \frac{ \delta }{\delta u } \wedge \left [ \frac{ \delta }{\delta w } \right ]_{x} + 2 \frac{ \delta }{\delta v } \wedge \left [ \frac{ \delta }{\delta w } \right ]_{x} \right ] dx 
\end{eqnarray}
Poisson bracket defined by bivector field $\hat{W}$ gives rise to the second Hamiltonian realization of Benney system 
\begin{eqnarray}
u_{t} = \{h^{\bullet } , u\}_{\bullet }\nonumber \\v_{t} = \{h^{\bullet } , v\}_{\bullet }\nonumber \\w_{t} = \{h^{\bullet } , w\}_{\bullet }\nonumber \\
\end{eqnarray}
with new Hamiltonian 
\begin{eqnarray}
h^{\bullet } = \frac{ 1 }{6 } \int ^{ + \infty }_{- \infty } (v^{2} + 2uw)dx 
\end{eqnarray}
Thus symmetry (\ref{eq:e144}) is closely related to bi-Hamiltonian realization of Benney hierarchy. \\
The same symmetry yields infinite sequence of conservation laws of Benney system. Namely one can construct sequence of integrals of motion by applying non-Noether symmetry (\ref{eq:e144}) to 
\begin{eqnarray}
J^{(0)} = \int ^{ + \infty }_{- \infty } wdx 
\end{eqnarray}
(the fact that $J^{(0)}$ is conserved can be verified by integrating third equation of Benney system). The sequence looks like 

\begin{eqnarray}
\label{eq:e148}
J^{(0)} = \int ^{ + \infty }_{- \infty } wdx \nonumber \\J^{(1)} = L_{E}J^{(0)} = 2 \int ^{ + \infty }_{- \infty } vdx \nonumber \\J^{(2)} = L_{E}J^{(1)} = (L_{E})^{2}J^{(0)} = 8 \int ^{ + \infty }_{- \infty } udx \nonumber \\J^{(3)} = L_{E}J^{(2)} = (L_{E})^{3}J^{(0)} = 12 \int ^{ + \infty }_{- \infty } (v^{2} + 2uw)dx \nonumber \\J^{(4)} = L_{E}J^{(3)} = (L_{E})^{4}J^{(0)} = \nonumber \\48 \int ^{ + \infty }_{- \infty } (2uw^{2} + 4uv + v^{2}w)dx \nonumber \\J^{(5)} = L_{E}J^{(4)} = (L_{E})^{5}J^{(0)} = \nonumber \\240 \int ^{ + \infty }_{- \infty } (4u^{2} + 8uvw + 2uw^{3} + 2v^{3} + v^{2}w^{2})dx \nonumber \\J^{(n)} = L_{E}J^{(n - 1)} = (L_{E})^{n}J^{(0)} 
\end{eqnarray}
So conservation laws and bi-Hamiltonian structure of Benney hierarchy are closely related to its symmetry, that can play important role in analysis of Benney system and other models that can be obtained from it by reduction. \\
Similarly to the case of nonlinear water wave system, proof of involutivity of these conservation laws is based on Theorem 8. Namely, 1-form $s$ defined by means of non-Noether symmetry (\ref{eq:e144}) as follows 
\begin{eqnarray}
E = W(s)
\end{eqnarray}
has property 
\begin{eqnarray}
[W[W(s),W](s)] = \frac{ 1 }{2 }[W(s)[W(s) ,W]] 
\end{eqnarray}
while conservation law 
\begin{eqnarray}
J = \int ^{ + \infty }_{- \infty } udx 
\end{eqnarray}
satisfies condition 
\begin{eqnarray}
W(L_{W(s)}dJ) = - [W(s),W](dJ) 
\end{eqnarray}
and thus according to Theorem 8 produces involutive family of conserved quantities. \\
\section{Conclusions}
The fact that many important integrable models, such as Korteweg-de Vries equation, nonlinear Schr\"{o}dinger equation, Broer-Kaup system, Benney system and Toda chain, possess non-Noether symmetries that can be effectively used in analysis of these models, inclines us to think that non-Noether symmetries can play essential role in theory of integrable systems and properties of this class of symmetries should be investigated further. The present review indicates that in many cases non-Noether symmetries lead to maximal involutive families of functionally independent conserved quantities and in this way ensure integrability of dynamical system. To determine involutivity of conservation laws in cases when it can not be checked by direct computations (for instance in direct way one can not check involutivity in many generic n-dimensional models like Toda chain and infinite dimensional models like KdV hierarchy) we propose analog of Yang-Baxter equation, that being satisfied by generator of symmetry, ensures involutivity of family of conserved quantities associated with this symmetry.\\
Another important feature of non-Noether symmetries is their relationship with several essential geometric concepts, emerging in theory of integrable systems, such as Fr\"{o}licher-Nijenhuis operators, Lax pairs, bi-Hamiltonian structures and bicomplexes. From one hand this relationship enlarges possible scope of applications of non-Noether symmetries in Hamiltonian dynamics and from another hand it indicates that existence of invariant Fr\"{o}licher-Nijenhuis operators, bi-Hamiltonian structures and bicomplexes in many cases can be considered as manifestation of hidden symmetries of dynamical system. \\
\section{Acknowledgements}
Author is grateful to George Jorjadze, Zakaria Giunashvili and Michael Maziashvili for constructive discussions and help. This work was supported by INTAS (00-00561).\\

\end{document}